%% file: isomain.tex
\documentclass[12pt,reqno]{amsart}
\usepackage{epsfig}

\newcommand{\bC}{{\mathbb C}}
\newcommand{\bR}{{\mathbb R}}

\newcommand{\cD}{\mathcal{D}}

\newcommand{\vV}{\mathbf{V}}
\newcommand{\vW}{\mathbf{W}}

\newcommand{\vD}{\mathbf{D}}
\newcommand{\ve}{\mathbf{e}}

\newcommand{\vr}{\mathbf{r}}

\newcommand{\vz}{\mathbf{z}}

\newcommand{\vv}{\mathbf{v}}

\newcommand{\vomega}{\boldsymbol{\omega}}
\newcommand{\rphi}{\phi}
\newcommand{\Rphi}{\Phi}
\newcommand{\bloch}{\varphi}
\newcommand{\baker}{\psi}
\newcommand{\vrphi}{\boldsymbol{\rphi}}
\newcommand{\vbloch}{\boldsymbol{\bloch}}
\newcommand{\vbaker}{\boldsymbol{\baker}}

\newcommand{\vphi}{\boldsymbol{\phi}}
\newcommand{\vpsi}{\boldsymbol{\psi}}

\newcommand{\vone}{\boldsymbol{1}}
\newcommand{\A}{\boldsymbol{\mathcal A}}
\newcommand{\vgamma}{{\boldsymbol{\gamma}}}

\newcommand{\interchange}{\iota}

\newcommand{\CC}{{\mathbb C}}

\newcommand{\R}{{\mathbb R}}

\def\widebar{\overline}

\def\ubar{\widebar{u}}

\def\realpart{\operatorname{\sf Re}}
\def\impart{\operatorname{\sf Im}}
\def\sign{\operatorname{sign}}

\newcommand{\di}{\partial}

\newcommand{\bel}[2]{\begin{equation}\label{#1} #2 \end{equation}}

\newcommand{\rd}{\mathrm{d}}            
\newcommand{\re}{\mathrm{e}}            
\newcommand{\ri}{\mathrm{i}}            

\newtheorem{theorem}{Theorem}
\numberwithin{theorem}{section}
\newtheorem{corollary}[theorem]{Corollary}

\newtheorem{lemma}[theorem]{Lemma}

\newtheorem{proposition}[theorem]{Proposition}
\newtheorem{prop}[theorem]{Proposition}

\theoremstyle{definition}

\begin{document}
\title[Finite-gap Solutions of the Vortex Filament Equation]{Finite-gap Solutions of the Vortex Filament Equation: Isoperiodic Deformations}
\author{A. Calini and T. Ivey}
\thanks{The authors were partially funded by NSF grants DMS-0204557 and DMS-0608587}
\address{Department of Mathematics, College of Charleston \\ Charleston SC 29424 USA}
\email{calinia@cofc.edu, iveyt@cofc.edu}

\date{\today }
\maketitle
\begin{abstract}
We study the topology of quasiperiodic solutions of the vortex filament
equation in a neighborhood of multiply covered circles. We construct these
solutions by means of a sequence of isoperiodic deformations, at each step
of which a real double point is ``unpinched" to produce a new pair of
branch points and therefore a solution of higher genus. We prove that
every step in this process corresponds to a cabling operation on the
previous curve, and we provide a labelling scheme that matches the
deformation data with the knot type of the resulting filament.
\end{abstract}

\input{isointro}

\input{isodeforms}

\input{isoexamples}

\input{isoperturb}

\input{isocabling}

\input{isoapp1}

\input{isogap}

\input{isobib}

\end{document}

%% file: isointro.tex
\section{Introduction}

In this sequel to \cite{CI05}, we continue our study of the role of integrability for periodic solutions of the Vortex Filament Equation (also known as Localized Induction Equation)
\begin{equation}
\label{VFE}
\frac{\partial \vgamma}{\partial t}= \frac{\partial \vgamma}{\partial x}\times \frac{\partial^2 \vgamma}{\partial x^2},
\end{equation}
a model of the self-induced dynamics of a vortex line in a Eulerian fluid,
described in terms of the evolution of the position vector $\vgamma(x,t)$
of a space curve parametrized by arclength $x$.

Hasimoto's transformation \cite{Ha}
\bel{Hmap}{
q = \tfrac12\, \kappa \exp\left({\ri \int \tau\, \rd s}\right),}
given in terms of the curvature $\kappa$ and torsion $\tau$ of $\vgamma$,
maps the Vortex Filament Equation (VFE) to the focussing cubic nonlinear
Schr\"odinger (NLS) equation
\begin{equation}
\label{NLS2}
\ri q_t + q_{xx} + 2 |q|^2 q = 0,
\end{equation}
with the NLS potential $q$ defined up to multiplication by an arbitrary
constant phase factor. The integrability of the NLS equation  \cite{FT, ZS}
implies that the VFE inherits many of the properties of the integrable
equation, including a family of global geometric invariants (conserved
quantities) \cite{LP}, a bihamiltonian structure \cite{Ca, LP, Sa, SY14},
and special solutions: solitons, finite-gap solutions, and their
homoclinic orbits \cite{CI1, CI05, CGS, Sy}.

Periodic boundary conditions for the VFE give rise to closed curves; of
those, the class of vortex filaments corresponding to periodic and
quasi-periodic finite-gap NLS potentials provides contains  striking
examples of curves exhibiting special geometric features (such as symmetry
and periodic planarity) and interesting topology. Our previous article
\cite{CI05} focused on  characterizing geometric properties of finite-gap
VFE solutions such as closure, symmetries, self-intersection, and
planarity in terms of the Floquet spectrum of associated finite-gap NLS
potentials. In contrast, the current work concerns the topological
information  contained in the algebro-geometric data of  a closed VFE
solution associated with a periodic finite-gap NLS potential.

Before describing our approach to this problem, we will briefly introduce some of the main tools and results used in the paper.

\noindent
{\bf The NLS linear system and the curve reconstruction formula.} The AKNS linear system for   \eqref{NLS2} consists of a pair of first-order linear systems \cite{FT}:
an eigenvalue problem
\begin{equation}
\label{AKNSx}
\mathcal{L}_1\vphi=\lambda \vphi,
\end{equation}
and an evolution equation
\begin{equation}
\label{AKNSt}
\vphi_t=\mathcal{L}_2\vphi
\end{equation}
 for the complex vector-valued eigenfunction $\vphi$. The solvability or
``zero curvature'' condition of the AKNS system is  the NLS equation
\eqref{NLS2}. Expressed in terms of the Pauli matrix
$\sigma_3 = \begin{pmatrix}1 & 0 \\ 0 & -1 \end{pmatrix}$,
$$
\mathcal{L}_1 =\ri \sigma_3 \frac{\partial}{\partial x}+  \begin{pmatrix}0 & q \\ - \bar{q} & 0 \end{pmatrix},
\qquad
\mathcal{L}_2 = \ri (|q|^2-2\lambda^2) \sigma_3 +
\begin{pmatrix} 0 & 2\ri \lambda q - q_x \\
 2\ri \lambda \bar{q} +\bar{q}_x & 0 \end{pmatrix}.
$$
The coefficients of the linear operators $\mathcal{L}_1$ and $\mathcal{L}_2$ depend on $x$ and $t$ through the complex-valued
function $q$, and on the {\em spectral parameter} $\lambda \in \bC$.

The inverse of the Hasimoto map (i.e.,  the reconstruction of a curve
given its curvature and torsion) is realized in terms of the solutions of
the AKNS system (equivalent to the Darboux equations for the natural frame
of the curve). Remarkably, the reconstruction of the evolving filament
requires taking no antiderivatives: given a fundamental matrix  solution
$\Phi(x,t;\lambda)$ of the AKNS system, such that $\Phi(0,0;\lambda)$ is a
fixed element of $SU(2)$, then the skew-hermitian matrix
\begin{equation}
\label{RECO}
\vgamma(x,t)=\left. \Phi^{-1}\frac{\rd \Phi}{\rd\lambda} \right|_{\lambda=0}
\end{equation}
satisfies the VFE \eqref{VFE}, and corresponds to $q$ via the Hasimoto map \cite{Pohlm, SY14}.
(We have identified $\frak{su}(2)$ with $\R^3$ via a fixed isometry, under which the Lie bracket corresponds to -2 times the
cross product.)
Formula \eqref{RECO},  known as the {\sl Sym-Pohlmeyer reconstruction
formula}, can also be evaluated at a nonzero real eigenvalue $\lambda=\Lambda_0$. The resulting curve
 $\vgamma$ still satisfies \eqref{VFE}, but with a potential that differs from what we
started with by the Galilean transformation $q(x,t) \mapsto \re^{\ri(a x-a^2 t)} q(x-2a t,t)$, $a=-2\Lambda_0$, which
preserves solutions of \eqref{NLS2}. Given a closed curve of length $L$,
the potential $q$ obtained by the Hasimoto map \eqref{Hmap} is not
necessarily $L$-periodic, but will be related to an $L$-periodic potential
by a Galilean transformation. The closed curve may then be recovered using
\eqref{RECO}, but evaluated at $\lambda=\Lambda_0$.

\noindent
{\bf The Floquet spectrum of a finite-gap solution.} The spectrum
associated with an $L$-periodic NLS potential $q(x)$ is defined in terms
of the {\sl Floquet discriminant }
$$
\Delta(q;\lambda)=\mathrm{Trace}(\Phi(x+L,t;\lambda)\Phi(x,t;\lambda)^{-1}),
$$
the trace of the transfer matrix across one period $L$, where $\Phi$ is a
fundamental matrix solution of the AKNS system.  The Floquet spectrum is
the set of complex  $\lambda$ values for which the eigenfunctions of the
AKNS system are bounded on the spatial domain:
$$
\sigma(q)=\left\{ \lambda\in \mathbb{C} \, | \, \Delta(q;\lambda )\in \mathbb{R}, -2\leq \Delta \leq 2\right\}.
$$
It can be shown that $\Delta$ is a conserved functional of the NLS time
evolution, and in fact a generating function of the constants of motion.
In particular, the spectrum of a given NLS potential is invariant under
the evolution.

\begin{figure}
\label{ngapspectrum}
\begin{center}
\includegraphics[width=.45\textwidth]{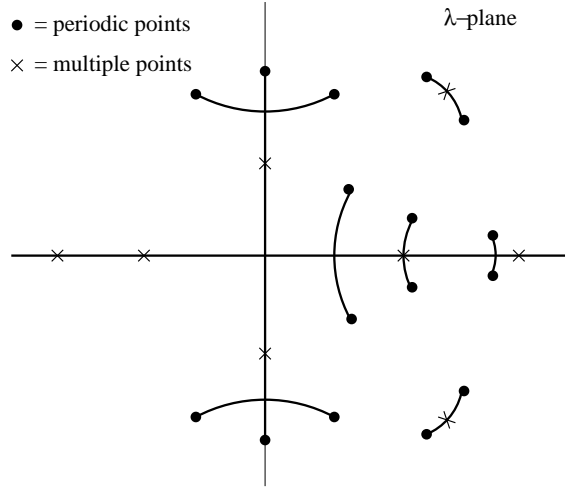}
\caption{The Floquet spectrum of a finite-gap NLS potential.}
\end{center}
\end{figure}
In Figure \ref{ngapspectrum}, we show a  typical spectrum for a finite-gap
potential, which has a finite number of complex bands of continuous
spectrum. Those points at which $\Delta=\pm2$ are divided into {\em simple
points} and {\em multiple points}, according to their multiplicity as
zeros of $\Delta^2-4$. (Finite-gap potentials are characterized by having
a finite number of simple points.) Multiple points are among the {\em
critical points} for $\Delta$; for example, double points satisfy
$\rd{\Delta}/\rd \lambda =0$,  but $\rd^2{\Delta}/\rd \lambda^2 \not= 0$.
Real double points are removable \cite{Schm}, i.e., at these
$\lambda$-values the transfer matrix has a pair of  linearly independent
eigenvectors.

Critical points where $\Delta$ does not achieve its maximum or minimum
value are embedded within bands of continuous spectrum. It can be easily
shown that $\rd{\Delta}/\rd \lambda =0$ at transverse intersections of
bands of continuous spectrum. Those critical points associated with spines
intersecting the real axis play an important role in the closure condition
of the reconstructed curve, as well as in our analysis.

\noindent
{\bf The quasi-momentum differential and the closure condition.}
At a
generic complex $\lambda$, the transfer matrix has a pair of distinct
eigenvalues $\rho^+(\lambda)$, $\rho^-(\lambda)$ known as {\sl Floquet
multipliers}. The relation between Floquet multipliers and Floquet
discriminant is given by
\[
\rho^\pm(\lambda)=\frac{\Delta(\lambda) \pm \sqrt{\Delta(\lambda)^2-4}}{2}.
\]
Thus, $\rho^\pm(\lambda)$ are branches of a holomorphic function $\rho$ which is well defined on a two-sheeted Riemann surface $\Sigma$,
whose projection $\pi: \Sigma \rightarrow \mathbb{C}$ is branched at the simple points. For $P \in \Sigma$, we will let
\[
\rho(P)=\re^{\ri L \Omega_1(P)},
\]
where the function $\Omega_1(P)$ is defined up to adding an integer multiple of $2\pi/L$. Its differential
\[
\rd \Omega_1=\frac{1}{\ri L}\ \rd \log \rho,
\]
known as the {\sl quasimomentum differential}, is a well-defined meromorphic differential on $\Sigma$.
Because $\rd \Omega_1$ changes by minus one from sheet to sheet, each pair of its zeros projects to a single $\lambda$-value,
and we will regard its zeros as being located in the complex plane.

For a given finite-gap NLS potential, the real zeros of $\rd \Omega_1$
(which are the real critical points) play a role in the following result
by Grinevich and Schmidt \cite{GSc}:

\bigskip
\noindent
{\bf Closure Conditions.} {\sl  A finite-gap VFE solution $\vgamma$ obtained
from the generalized Sym-Pohlmeyer reconstruction formula \eqref{RECO} at
$\lambda=\Lambda_0$ is smoothly closed if the reconstruction point
$\Lambda_0$ is
(i) a real double point and
(ii) a zero of the quasimomentum differential.
}

\bigskip
\noindent
(See also \cite{CI05} for a derivation of the closure conditions from the explicit formulas for finite-gap VFE solutions.)
\bigskip

Closure conditions must of course be satisfied if one is interested in
establishing connections between the knot types of closed finite-gap VFE
solutions and the spectra of the associated NLS potentials. However, such
conditions turn out to be difficult to compute, as they involve solving
implicit equations written in terms of hyperelliptic integrals.

The central idea of this paper is to approach the problem of constructing
closed finite-gap solutions of the VFE by starting with an already closed
curve of particularly simple type (namely a multiply covered circle) and
deforming its associated spectrum in such a way that both periodicity of
the corresponding NLS potential and closure of the curve are preserved. By
iterating similar isoperiodic deformation steps, we construct a
neighborhood of the initial curve that consists of closed finite-gap
solution of increasing complexity. (Indeed, at each step of the
deformation, a real double point is ``opened up" into a new pair of branch
points, thus increasing the genus of the Riemann surface by one.)

A beautiful consequence of this multi-step construction is a {\sl
labelling scheme} that matches the deformation data with the knot type of
the resulting filament. Our main result,  the Cabling Theorem (Theorem
\ref{cabletheorem_formal}) is that every step in this process is, from the
topological point of view, a cabling operation on the previous curve, with
the cabling type encoded in which real double points are selected to be
deformed into a new pair of branch points. A simplified statement of this
result is:

\medskip
\noindent
{\bf Cabling Theorem.}\quad
{\sl Given relatively prime integers $n, m_1, \ldots, m_K$ such that
$|m_k|>n>1$, let $g_k = \gcd(n,m_1, \ldots, m_k)$ for $1\le k \le K$.  Then there exist
a sequence of deformations (i.e., one-parameter families) of finite-gap potentials $q^{(k)}(x;\epsilon)$ of fixed period $L$,
 and complex numbers $\Lambda_0^{(k)}(\epsilon)$,
which are both analytic in $\epsilon$, such that
\newline
(1) $q^{(k)}$ is of genus $k$ when $\epsilon \ne 0$, $q^{(k)}(x;0)=q^{(k-1)}(x,\epsilon_{k-1})$
for some $\epsilon_{k-1}>0$ when $k>1$, and $q^{(1)}(x;0)$ is constant;
\newline
(2) the filament $\vgamma^{(k)}(x;\epsilon)$ that is constructed from $q^{(k)}(x;\epsilon)$ using
the Sym-Pohlmeyer formula evaluated at $\Lambda_0^{(k)}(\epsilon)$ is
closed of length $n\pi/g_k$ and is, for $\epsilon$ sufficiently small, a
$(g_{k-1}/g_k, m_k/g_k)$-cable
about $\vgamma^{(k)}(x;\epsilon_{k-1})$.}

\bigskip

The full statement of Theorem \ref{cabletheorem_formal} explains how these
deformations arise from isoperiodic deformations of the spectrum, and how
the data $[n, m_1, \ldots, m_K]$ determine the selection of double points
to be opened up. From an argument in the proof of Theorem
\ref{cabletheorem_formal}, we deduce as a bonus that the knot types of
the finite-gap filaments so constructed are constant under the VFE
evolution (see Corollary \ref{timecorollary}), ultimately confirming and
justifying the use of the Floquet spectrum as a tool for classifying the
knot types of closed curves in an appropriate neighborhood of multiply
covered circles. (In fact, such curves are approximated by finite-gap
solutions that are dense in the space of periodic solutions of the VFE
\cite{Gri}.)

The proof of the main result brings together a variety of tools from the
periodic theory of integrable systems and the perturbation theory of
ordinary differential equations. We mention them below while giving a
brief description of the paper contents and organization.

Section 2 contains the motivation of this work and the framework of our
approach. After introducing Grinevich and Schmidt's isoperiodic
deformation system,  we define a special type of closure-preserving
isoperiodic deformation that reverses ``pinching" of the two ends of a
spine of spectrum into a real double point (homotopic deformations), and
show that the solution to such a deformation is analytic in the
deformation parameter.  We then proceed to deform off the spectrum of a
modulationally unstable plane wave solution (corresponding to a
multiply-covered circle solution of the VFE), and compute the spatial
frequencies of the resulting finite-gap solutions. Examples of the curves
resulting from successive homotopic deformations, and a labelling scheme
for their knot types are presented in Section 3. Section 4 makes use of
the completeness of the family of squared eigenfunctions for the AKNS
system to characterize, to first order, the perturbations of the potential
$q$ associated with homotopic deformations. The Cabling Theorem and its
proof are the contents of Section 5: the proof is a combination of
explicit perturbative computations involving the squared eigenfunctions,
a topological argument based on White's formula for self-linking, and a
careful analysis of the argument of the first order correction to the initial
potential, which determines the cabling phenomenon and the cable type. The
two appendices (Sections 6 and 7) contain a statement of the completeness
theorem for squared eigenfunctions and related useful results, and a proof
of the analytic dependence of the potential $q$ on the deformation parameter.

%% file: isodeforms.tex
\section{Isoperiodic Deformations}\label{isodeforms}

Inspecting the formula \eqref{finiteq} for a finite-gap NLS solution $q(x,t)$ shows
that $q$ is periodic in $x$ if and only if the components of the frequency
vector $\vV \in \bR^g$ and a real scalar $E$ are rationally related.  These data
are determined by a choice of $g+1$ pairs of conjugate branch points in the
complex plane.  Furthermore, because $E$ changes additively when the branch points
are translated in the real direction, construction of a periodic solution depends
on being able select the components $V_j$ of the frequency vector.  We now describe
a scheme for deforming the spectrum of a multiply-covered circle which produces arbitrary
rational values for these components.

Grinevich and Schmidt \cite{GrS0} developed a method for deforming the branch points
in a way that preserves the components $V_k$ of the
frequency vector.  Such isoperiodic deformations of the spectrum were
first introduced by Krichever \cite{Kriso} in connection with topological
quantum field theory, and are naturally related to the Whitham equations
in the work of Flaschka, Forest and McLaughlin \cite{FFMcL}. Although the
zeros $\alpha_1, \ldots , \alpha_{g+1}$ of the quasimomentum differential $\rd\Omega_1$
are dependent on the branch points  $\lambda_1, \ldots, \lambda_{2g+2}$,
when we incorporate these $\alpha_k$ as dependent variables
the isoperiodic deformation becomes a system of ordinary
differential equations with rational right-hand sides:
\begin{equation}
\begin{split}
\frac{\rd \lambda_j}{\rd \xi} &=
  -\sum_{k=1}^{g+1}\frac{c_k}{\lambda_j-\alpha_k}
  \qquad 1 \le j \le 2g+2,
  \\
\frac{\rd \alpha_k}{\rd \xi} &=
    \sum_{\ell\ne k}\frac{c_k+c_\ell}{\alpha_\ell -\alpha_k}
    -\frac{1}{2}\sum_{j=1}^{2g+2}\frac{c_k}{\lambda_j-\alpha_k},
    \qquad 1 \le k,\ell \le g+1.
\end{split}
\label{isoperiodicsystem}
\end{equation}
Here, $c_1, \ldots, c_{g+1}$ are controls, i.e., arbitrary functions of
the real deformation parameter $\xi$.  In the case of finite-gap NLS solutions,
the $\lambda_j$ and $\alpha_k$ are roots of a real polynomial, and it is
easily checked that complex conjugacy relationships among these roots (e.g., $\lambda_{2j+2} = \overline{\lambda_{2j+1}}$,
$\alpha_2 = \overline{\alpha_1}$, $\alpha_3 = \overline{\alpha_3}$) are preserved by these deformations,
provided that the controls $c_k$ have the same conjugacy relationships
as the $\alpha_k$.

The change in the value of the quasimomentum $\Omega_1$ at one of the
$\alpha_k$ under this deformation is given by
\[
\frac{\rd}{\rd \xi} \Omega_1(\alpha_k)=
c_k \left. \left( \frac{1}{\lambda - \alpha_k} \frac{\rd \Omega_1}{\rd \lambda}   \right) \right|_{\lambda=\alpha_k}.
\]
Thus, we may preserve the value of $\Omega_1(\alpha_k)$ simply by setting $c_k$
to zero.  In particular, if $\alpha_k$ is real and the value of $\Omega_1(\alpha_k)$ is such that the Sym-Pohlmeyer
reconstruction formula \eqref{RECO} yields a closed curve at $\Lambda_0=\lambda_k$,
then by choosing the control $c_k=0$ the isoperiodic deformation will
produce a closed curve for every $\xi$. We will refer to this
specialization of isoperiodic deformations as a
{\em homotopic deformation}, since it generates a homotopy through the family of smooth
maps of the circle into $\mathbb{R}^3$.

Instances of homotopic deformation have been observed before.
David Singer and the second author \cite{IS} constructed one-parameter
families of closed elastic rod centerlines (which correspond to genus one finite-gap
NLS solutions under the Hasimoto map) in the form of torus knots, terminating
in multiply-covered circles at either end of the deformation.
We can try to generate this deformation using system \eqref{isoperiodicsystem}
with $g=1$.
Assume that, say, $\alpha_2$ is the real critical point that yields
a closed curve.  (Necessarily, the other critical point $\alpha_1$ must
be real.) Then choosing $c_2=0$ and $c_1=1$ will reproduce
part of this homotopic deformation.  The `circular' end of the deformation occurs
when $\alpha_1$ and one of the complex conjugate pairs of branch points
(say, $\lambda_1$ and $\overline{\lambda_1}$)
limit to the same real value as $\xi$ decreases to some finite time $T_1$.
Note that this is a singularity of the isoperiodic deformation system, as the right-hand side
of \eqref{isoperiodicsystem} blows up as $\lambda_1 \rightarrow \alpha_1$.
In the next subsection, we will examine
this type of singularity for \eqref{isoperiodicsystem} in more detail.

The genus one solution of \eqref{isoperiodicsystem} discussed above also reaches a singularity
in finite forward time, when the two $\alpha$'s collide.  In our previous
paper \cite{CI05} we showed that this other kind of singularity
is associated with the elastic rod centerline becoming an Euler elastic curve.
Presumably, the solution may be continued smoothly through the singularity,
although we have not investigated this question.

\subsection{Pinch-Type Singularities}
We will say that a solution of \eqref{isoperiodicsystem} (in arbitrary genus)
has a {\em pinch-type singularity} when exactly two complex conjugate branch points
and exactly one real critical point $\alpha_k$ approach the same real value.  (The reason
for the name is that bringing two branch points together collapses a homotopy
cycle on the associated Riemann surface, hence pinching one handle on
a $g$-handled torus.)  We will limit our attention to the case where
only the control associated to $\alpha_k$ is nonzero.

Because the system \eqref{isoperiodicsystem} is
invariant under permuting the indices  on the branch points, and
invariant under simultaneously permuting the indices on the critical points
and the controls, we can without loss of generality
let $\lambda_1$ and $\lambda_2 = \overline{\lambda_1}$ be the
colliding branch points, and $\alpha_1$ the critical point, with $c_1 = 1$ being
the only nonzero control.  Then the system takes the form
\begin{equation}
\begin{split}
\frac{\rd \lambda_j}{\rd \xi} &=
  -\frac{1}{\lambda_j-\alpha_1},
  \qquad 1 \le j \le 2g+2,
  \\
\frac{\rd \alpha_1}{\rd \xi} &=
    \sum_{k=2}^{g+1}\frac{1}{\alpha_k -\alpha_1}
    -\frac{1}{2}\sum_{j=1}^{2g+2}\frac{1}{\lambda_j-\alpha_1},
    \\
\frac{\rd \alpha_k}{\rd \xi} &=
    \frac{1}{\alpha_1 -\alpha_k},
     \qquad 2 \le k \le g+1.
\end{split}
\label{isoperiodicsingle}
\end{equation}
In particular, because $\lambda_2 = \overline{\lambda_1}$,
$$\frac{\rd (\lambda_1 - \alpha_1)}{\rd \xi}
= -\frac{1}{2}\left(\frac{1}{\lambda_1 - \alpha_1}-\frac{1}{\overline{\lambda_1}-\alpha_1}\right)
+\frac{1}{2}\sum_{j=3}^{2g+2}\frac{1}{\lambda_j-\alpha_1} - \sum_{k=2}^{g+1}\frac{1}{\alpha_k-\alpha_1},$$
showing that, if $\lambda_1 - \alpha_1$ approaches zero as $\xi \searrow 0$, and the other differences $\lambda_j-\alpha_1$
and $\alpha_k - \alpha_1$ have nonzero limits, then $\impart(\lambda_1-\alpha_1)$ should approach zero like $\sqrt{2\xi}$.

\begin{prop}\label{hardprop}
When the change of variable $\xi = \frac12 t^2$ is made in \eqref{isoperiodicsingle},
the resulting system has a solution which is analytic at $t=0$ and satisfies
\begin{align*}
\alpha_k &= \alpha^0_k + O(t^2), \qquad 1 \le k \le g+1\\
\lambda_1 &= \alpha^0_1 + \ri t  + O(t^2),\\
\lambda_2 &= \overline{\lambda_1},\\
\lambda_j &= \lambda^0_j + O(t^2), \qquad 3 \le j \le 2g+2,
\end{align*}
where $\alpha^0_1$ is real, and for $k>1$ and $j>2$, the values $\alpha^0_k,\lambda^0_j$ are distinct from $\alpha^0_1$.
\end{prop}
\begin{proof}
Let $\lambda_1 - \alpha_1 = x+\ri y$ and $z=\realpart(\lambda_1)$.  Then, because $\alpha_1$ is real,
$$\dfrac{\rd z}{\rd \xi} = -\realpart\left(\dfrac1{\lambda_1-\alpha_1}\right) =\dfrac{-x}{x^2+y^2}.$$
Because the remaining $\alpha$'s are real, and the $\lambda_j$ are in complex conjugate pairs (with $\lambda_2 = \overline{\lambda_1}$),
\begin{align*}
\dfrac{\rd y}{\rd \xi} &=-\impart\left(\dfrac1{\lambda_1-\alpha_1}+\sum_{k>1} \dfrac{1}{\alpha_k-\alpha_1} - \frac12\sum_j \dfrac{1}{\lambda_j-\alpha_1}\right)
\\
&= -\frac1{2}\impart\left(\dfrac{1}{\lambda_1-\alpha_1} - \dfrac1{\overline{\lambda_1}-\alpha_1}\right)=\dfrac{y}{x^2+y^2}.
\end{align*}
and
$$\dfrac{\rd x}{\rd \xi} =-\impart\left(\dfrac1{\lambda_1-\alpha_1}+\sum_{k>1} \dfrac{1}{\alpha_k-\alpha_1} - \frac12\sum_j \dfrac{1}{\lambda_j-\alpha_1}\right)
= \frac12\sum_{j>2}\dfrac{1}{\lambda_j-\alpha_1} - \sum_{k>1} \dfrac{1}{\alpha_k-\alpha_1} := N.$$
Note that the quantity $N$ is, by assumption, a nonzero analytic function of its arguments at their initial values
$\lambda_j = \lambda^0_j$, $\alpha_1 = \alpha^0_1$, and $\alpha_k = \alpha^0_k$, for $j>2$ and $k>1$.

We now change to $y$ as independent variable, and introduce the new dependent variables
$$\widetilde x = x/y, \quad \widetilde z = (z-\alpha^1_0)/y, \quad \widetilde \lambda_j = (\lambda_j - \lambda^0_j)/y,
\quad \widetilde \alpha_k = (\alpha_k - \alpha^0_k)/y.$$
Of course, if the conclusions of the proposition hold, then these new variables will be analytic functions
of $y$ that vanish to at least order one when $y=0$.  In terms of these new dependent and independent
variables, the system becomes
\begin{align*}
y \dfrac{\rd \widetilde x}{\rd y} &= - \widetilde x + y(1+\widetilde x^2)N \\
y \dfrac{\rd\widetilde z}{\rd y} &= - \widetilde x - \widetilde z, \\
y \dfrac{\rd\widetilde \lambda_j}{\rd y} &= -\widetilde \lambda_j
- \dfrac{y(1+\widetilde x^2)}{\lambda^0_j - \alpha^0_1 + y(\widetilde x - \widetilde z +\widetilde \lambda_j)},\\
y \dfrac{\rd\widetilde \alpha_k}{\rd y} &= -\widetilde \alpha_k
-  \dfrac{y(1+\widetilde x^2)}{\alpha^0_k - \alpha^0_1 + y(\widetilde x - \widetilde z  +\widetilde \alpha_k)}.\\
\end{align*}
This system has a Briot-Bouquet singularity at the origin.  Linearizing the right-hand sides at
$\widetilde x=\widetilde z = \widetilde \lambda_j = \widetilde \alpha_k=0$ gives a lower-triangular coefficient
matrix all of whose eigenvalues are equal to $-1$.  Thus, there is a unique solution all of whose
components vanish when $y=0$ and which is analytic in $y$ near $y=0$ (see, for example,
p. 21 in \cite{Iwano} or Theorem 59 in \cite{Zubov}\footnote{%
In the sources cited, the results are given for Briot-Bouquet systems
$z \rd y_i /\rd z = F_i(z,y_1,\ldots,y_n)$
where the linearization has $k$ positive eigenvalues,
the resulting solution is a double power series in $z$ and $\ln z$ that
converges for $|z|$ sufficiently small in a sector about the origin
in the complex plane, and the solution depends on $k$ constants.
It is implicit in the statements of these results that, when $k=0$, the solution is
unique and analytic in $z$.  In that case, the existence/uniqueness
result can also be proved in a more elementary way using
the method of majorants.}%
).
Consequently, $x,z,\lambda_j$ and $\alpha_k$ are analytic functions of $y$
which satisfy
\begin{equation}\label{owhybehaviour}
x = O(y^2),\quad z = \alpha^0_1 + O(y^2), \quad \lambda_j = \lambda^0_j + O(y^2), \quad \alpha_k = \alpha^0_k + O(y^2).
\end{equation}

Because $\rd \xi /\rd y = y(1+\widetilde x^2)$, it follows that $\xi$ is an analytic function of $y^2$, and we can arrange
that $\xi = 0$ when $y=0$.  Then $\xi = \frac12 y^2 + O(y^4)$.  Thus, we can define an analytic function $t$ of
$y$ such that $\xi = \frac12 t^2$ and $t = y + O(y^2)$.  Clearly, this function is invertible for $|y|$ sufficiently
small, so we may replace $y$ by $t$ in \eqref{owhybehaviour}, and assert that $y$ is an analytic function
of $t$ satisfying $y = t + O(t^2)$.  Rewriting these equations in terms of the original
variables $\lambda_1$ and $\alpha_1$ completes the proof.
\end{proof}


Note that the variable $t$ in Proposition \ref{hardprop} will be replaced by the more commonly-used
deformation parameter $\epsilon$ in later sections of this paper.

\subsection{Deforming the Multiply-Covered Circle}
Our main idea is to use the isoperiodic deformations provided by Proposition \ref{hardprop}
to create new, higher-genus NLS solutions while maintaining the closure of
the filament (in effect, reversing the collapse of branch points
described above in the genus one case).  By continuity, the initial value $\alpha^0_1$ used must be a
real point of the discrete spectrum (hence, a real double point) of the lower-genus NLS solution.
So, when we carry out several deformations in succession, we must keep track of the
continuously changing positions of those double points that we want to split later.
We can do this by incorporating them as extra variables in \eqref{isoperiodicsingle},
consisting of one critical point and a pair of branch points that remain equal
as the deformation progresses.  (Note that, for example, the equality
$\lambda_3=\lambda_4=\alpha_2$ is preserved under the deformation
\eqref{isoperiodicsingle}.)

We will begin the multi-step deformation process by selecting $g$ double
points from the spectrum of a multiply-covered circle
to be the initial values for $\alpha_1, \ldots, \alpha_g$,
and begin the deformation by setting $\lambda_{2k}=\lambda_{2k-1}=\alpha_k$ for $1\le k \le g$.
(There will be two addition branch points with initial values $\lambda_{2g+1}=\ri$ and $\lambda_{2g+2}=-\ri$,
and an additional $\alpha_{g+1}$ initially equal to zero.)  The first deformation, which
produces a genus one (elastic rod) solution, will be continued up to a small value of
the deformation parameter.  Then, $\alpha_2$ replaces $\alpha_1$ as the initial value for
the next double point to be split, and so on until all $g$ double points have been split.
Note that we will assume that the amplitude  of each deformation step (i.e., the maximum value of $\xi$ attained)
is sufficiently small so that singularities are avoided.  (Consequently, all critical points
remain real.)  As we will see in later sections,
it is also necessary to assume that the amplitude of each step is small relative to the last one, in order
to be able to predict the topological type of the resulting filament.

We will be able to calculate the frequencies $V_k$ for the genus $g$ NLS solution
by using the fact that the frequencies are unchanged by isoperiodic deformations,
and are determined by the integrals of certain holomorphic differentials
around cycles on the Riemann surface.  These integrals can, in turn, be
calculated by following the sequence of deformations back to the multiply-covered
circle, and doing residue calculations using the initial positions of the double points.

We begin this calculation with the spectrum of the multiply-covered circle.
Using the Hasimoto map \eqref{Hmap}, the potential $q(x)\equiv 1$ corresponds to a circle
with curvature $\kappa=2$, radius $1/2$, and length $\pi$.  A fundamental
matrix solution for the spatial part of the NLS linear system
\begin{equation}\label{bakerform}
\frac{\rd \vphi}{\rd x} = \begin{bmatrix} -\ri \lambda & \ri q \\ \ri \overline{q} & \ri \lambda \end{bmatrix}
\vphi
\end{equation}
with $q=1$ is given by
$$\Phi(x;\lambda) =
\begin{bmatrix}
 \cos(\omega x)-\frac{i\lambda}{\omega}\sin(\omega x) & \frac i{\omega} \sin(\omega x)\\
 \frac i{\omega} \sin(\omega x) & \cos(\omega x)+\frac{i\lambda}{\omega}\sin(\omega x)
\end{bmatrix}, \qquad \omega = \sqrt{1+\lambda^2}.
$$
The periodic spectrum for the singly-covered circle, computed relative to
period $L=\pi$, consists of two simple points $\lambda = \pm\ri$, and
infinitely many real double points given by $\lambda = \pm\sqrt{m^2-1}$,
$m=1,2,3,\ldots$.  (In fact, the origin is a point of order four.)
For the $n$-times-covered circle we compute the periodic spectrum
relative to period $L=n\pi$, and this consists of the same
simple points, but with double points given by
\begin{equation}\label{circlepoints}
\lambda = \pm \sqrt{(m/n)^2-1}, \qquad m=1,2,3,\ldots
\end{equation}
Note that the origin is the only point of the periodic spectrum
that can be used as the value $\Lambda_0$ producing a closed
curve via the Sym-Pohlmeyer reconstruction formula \eqref{RECO}.  Thus, zero will
be one of the initial values for the critical points in the first
step of the deformation process, and the corresponding control
will always be set to zero in order to maintain closure; we will
use $\Lambda_0$ to denote this `reconstruction point', which will
move along the real axis during the deformation steps.

Let $\beta_1 < \beta_2 < \ldots <\beta_g$
denote the choice of double points of
the spectrum of the $n$-times covered circle to be opened up.  (For the following
calculation, it is not relevant in what order they are opened up.) We will
now determine the relationship between these double points and the periods
of the genus $g$ solution obtained once all $g$ points have been opened up.
According to the construction for finite-gap solutions set forth in
\cite{BBEIM} and \cite{CI05}, the frequency vector $[V_1,\ldots, V_g]$ is
$4\pi \ri$ times the
first column of the inverse of the matrix $A$ defined by
$$A^j_k = \int_{a_k} \frac{\lambda^{g-j} \rd\lambda}{\zeta},$$
where $\zeta$ is related to $\lambda$ by the defining equation
of the hyperelliptic curve,
$$\zeta^2 = \prod_{j=1}^{2g+2}(\lambda-\lambda_j),$$
and each cycle $a_k$ circles around (in a clockwise fashion, on
the upper sheet\footnote{%
Along the real $\lambda$ axis, we initially label as the upper sheet that containing
the point $\infty_+$, where $\lambda^{g+1}/\zeta$ tends to $+1$ as $\lambda\to +\infty$,
but as one proceeds from right to left along the real axis in the $\lambda$-plane, the roles of upper sheet
and lower sheet are exchanged along branch cuts that run between each branch point
and its conjugate, parallel to the imaginary axis.}%
) the pair of complex conjugate branch points into which $\beta_k$ has been split
(see Figure \ref{althbasis}).

First, consider the case when all $\beta$'s are negative.  Then, as we run the deformation
steps backward, the cycles $a_1,\ldots,a_g$ become loops around $\beta_1,\ldots,\beta_g$,
and the only remaining branch points are $\pm \ri$.  The denominator $\zeta$ in the
integrand limits to $-\sqrt{\lambda^2+1}\,\prod_{j=1}^g (\lambda-\beta_j)$ along the $a$-cycles, taking the square root as positive along the real axis.
A residue calculation then gives
$$\lim A^j_k = N^j_k = \frac{2\pi \ri \beta_k^{g-j}}{\sqrt{\beta_k^2+1}\,\displaystyle\prod_{\ell\ne k} (\beta_k - \beta_\ell)}.$$
(Note that the minus sign in $\zeta$ is offset by the clockwise orientation of the $a$-cycles.)
We may factor this matrix as
\begin{equation}\label{Nmatrixdef}
N = 2\pi \ri \begin{bmatrix}
\beta_1^{g-1} & \ldots & \beta_g^{g-1} \\
\beta_1^{g-2} & \ldots & \beta_g^{g-2} \\
& \ldots & \\
1 & \ldots & 1 \end{bmatrix}
\begin{bmatrix}
1/D_1 & 0 & \ldots & 0 \\
0 & 1/D_2 & \ldots & 0 \\
& & \ldots & \\
0 & 0 & \ldots & 1/D_g
\end{bmatrix},\qquad D_k = \sqrt{\beta_k^2+1}\,\displaystyle\prod_{\ell\ne k} (\beta_k - \beta_\ell).
\end{equation}
Its inverse is
$$N^{-1} = \frac1{2\pi \ri}
\begin{bmatrix}
D_1 & 0 & \ldots & 0 \\
0 & D_2 & \ldots & 0 \\
& & \ldots & \\
0 & 0 & \ldots & D_g
\end{bmatrix}
\frac{1}{\prod_{j<k}(\beta_j - \beta_k)}
\begin{bmatrix}
\prod_{j<k; j,k \ne 1} (\beta_j - \beta_k) & \ldots \\
(-1)\prod_{j<k; j,k \ne 2} (\beta_j - \beta_k) & \ldots \\
(-1)^2 \prod_{j<k; j,k \ne 3} (\beta_j - \beta_k) & \ldots \\
 \ldots & \ldots \\
 \end{bmatrix}.
$$
(Only the first column is needed for the matrix on the right.) Thus,
\begin{equation}\label{freqleft}
V_j = 2\sqrt{\beta_j^2+1},\qquad 1\le j \le g.
\end{equation}

Next, suppose that $\beta_1 < \ldots < \beta_K < 0 < \beta_{K+1} < \ldots < \beta_g$.  Because
the branch cut between $\ri$ and $-\ri$, the roles of upper sheet and lower sheet are
switched an extra time, and the residue calculation gives
$$\lim A_k = \left\{\begin{aligned}
& N_k, \quad &  1\le k & \le K,\\
-& N_k, \quad & K+1 \le k &\le g,
\end{aligned}
\right.
$$
where $N_k$ is the $k$th column of the matrix defined by \eqref{Nmatrixdef} and
$A_k$ is the $k$th column of matrix $A$.
Solving for $\lim A^{-1}$ gives the following frequency formulas:
\begin{equation}\label{freqgen}
V_j= \left\{\begin{aligned}
2&\sqrt{\beta_j^2+1}, \quad & 1\le j &\le K,\\
-2&\sqrt{\beta_j^2 +1},\quad & K+1 \le j \le g
\end{aligned}
\right.
\end{equation}

%% file: isoexamples.tex
\section{Examples of cabling operations}
Comparing the frequency formulas \eqref{freqleft} and \eqref{freqgen}
with formula \eqref{circlepoints} for the spectrum of the multiply-covered
circle shows that when the double points $\beta_j$ are chosen as
members of that spectrum, the frequencies attained by the
deformations have rational values.  Accordingly, we will introduce
a notation for these deformations that allows these values to be
read off easily.

We will use the notation
$$[n; m_1, \ldots, m_g], \qquad |m_j|> n$$
to indicate the result of $g$ successive homotopic deformations of
the $n$-times covered circle, opening up the real double points
whose starting position is $\beta_j = -\sign(m_j) \sqrt{(m_j/n)^2-1}$
in the order in which the $m_j$ appear in the square brackets.\footnote{However, it matters
in which order the deformations are done, for it is easily checked that
if $\vv_j$ is the vector field on $\CC^{3g+3}$ defined by setting $c_j=1$ and
all other controls zero in the system \eqref{isoperiodicsystem}, then the
vector fields $\vv_j$ and $\vv_k$ do not commute when $j\ne k$.}
(The minus sign is incorporated here to compensate for the sign in \eqref{freqgen}.)
As mentioned before, we will assume that the amplitude of
each deformation step is small relative to that of the previous step.
We will also assume that the numbers in square brackets $m_i$ are relatively prime, so that
the double points selected do not all belong to the spectrum of a multiply-covered
circle for a smaller value of $n$.
In this notation, it is easy to see that $[n;m]$ gives frequency $V=m/n$,
$[n;m_1,m_2]$ gives frequency vector $[m_1/n, m_2/n]$, and so on.
These frequencies determine the length of the corresponding filament as follows:

The finite-gap NLS solution $q(x,t)$ takes the form
$$q(x,t) = A \exp(-\ri E x + \ri N t)\theta(\ri\vV x + \ri \vW t - \vD+\vr)/\theta(\ri \vV x + \ri \vW t - \vD),$$
where $A,E,N$ are constants, $\vV$ is the frequency vector, $\vD$ and $\vr$ are constant vectors in $\CC^g$,
and $\theta$ is a Riemann theta function with periods that are $2\pi\ri$ times an integer vector.
(For more details, see \S\ref{isogap}.)
The phase factor $\exp(-\ri E x)$ may be removed by an appropriate gauge transformation of NLS,
and then, assuming the components of $\vV$ are rational, the period of the potential $q$ is
the least common (integer) multiple of the periods $2\pi/V_j$.
The Baker-Akheizer functions, quadratic products of which are used to reconstruct the Frenet frame of the filament, take
the form
$$\psi_i = C_i(P) \exp(\ri x \Omega_1(P) + \ri t \Omega_2(P) ) \theta(\ri\vV x + \ri \vW t+\vD_i(P) )/\theta(\ri \vV x + \ri \vW t-\vD),
\qquad i=1,2,$$
where $P$ a point on the Riemann surface lying over the reconstruction point $\Lambda_0$, $\Omega_{1,2}$ are certain Abelian integrals on $\Sigma$,
and $C_{1,2}$ and $\vD_{1,2}$ are constants that depend on $P$.
Under the homotopic deformation
process we have described above, the value $\Omega_1(P)$ is preserved, so can be obtained
by calculating its limit as we run the deformation process backwards
to the multiply-covered circle.  Again, we first calculate this assuming that
all the limiting values $\beta_j$ of the double points are negative.  Then, because
$$\rd\Omega_1 = \dfrac{\displaystyle\prod_{j=1}^{g+1} (\lambda - \alpha_j)}{\zeta} \rd \lambda,$$
the limiting value of this differential, on the upper sheet above the origin, is
$$\lim\, \rd \Omega_1 = \dfrac{\lambda\, \rd \lambda}{\sqrt{\lambda^2 + 1}}.$$
Integrating from the limit $-\ri$ of the basepoint for $\Omega_1$
(which is by convention the point in the lower half-plane not enclosed by the basis $a$-cycles)
to the origin, which is the limit of the reconstruction point, gives
$\Omega_1(P) = 1$.
Next, assuming that the least positive double point is $\beta_{K+1}$,
$$\lim\, \rd\Omega_1 =(-1)^{g-K} \dfrac{\lambda \rd \lambda}{\sqrt{\lambda^2 + 1}}.$$
Integrating this from the basepoint to the origin gives $\Omega_1(P) = (-1)^{g-K}$.
Thus, the factor $\exp(2\ri x \Omega_1(P))$, which occurs in quadratic products
of Baker functions, has period $\pi$.  Since this factor is multiplied the theta functions,
the period of the Frenet frame is the least common multiple of
$\pi$ and the numbers $2\pi/V_j$ for $1\le j \le g$.  Because the reconstruction
point is chosen so that the integral of the Frenet unit tangent vector is zero,
this period is also the length of the filament.

\begin{figure}
\centering
\includegraphics[width=3.5in]{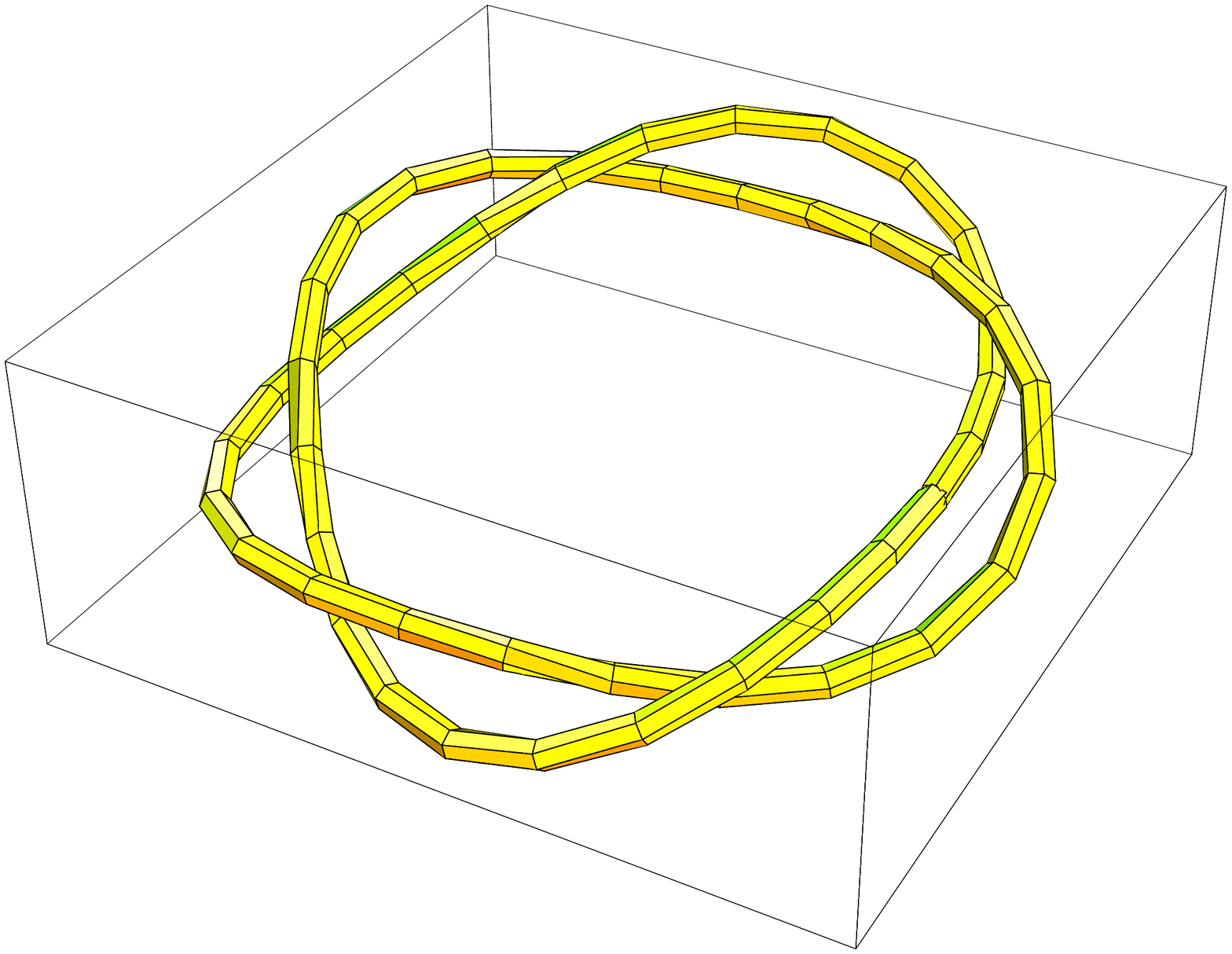}
\quad\begin{minipage}[b]{2.5in}\vspace{.5in}
\includegraphics[width=2.5in]{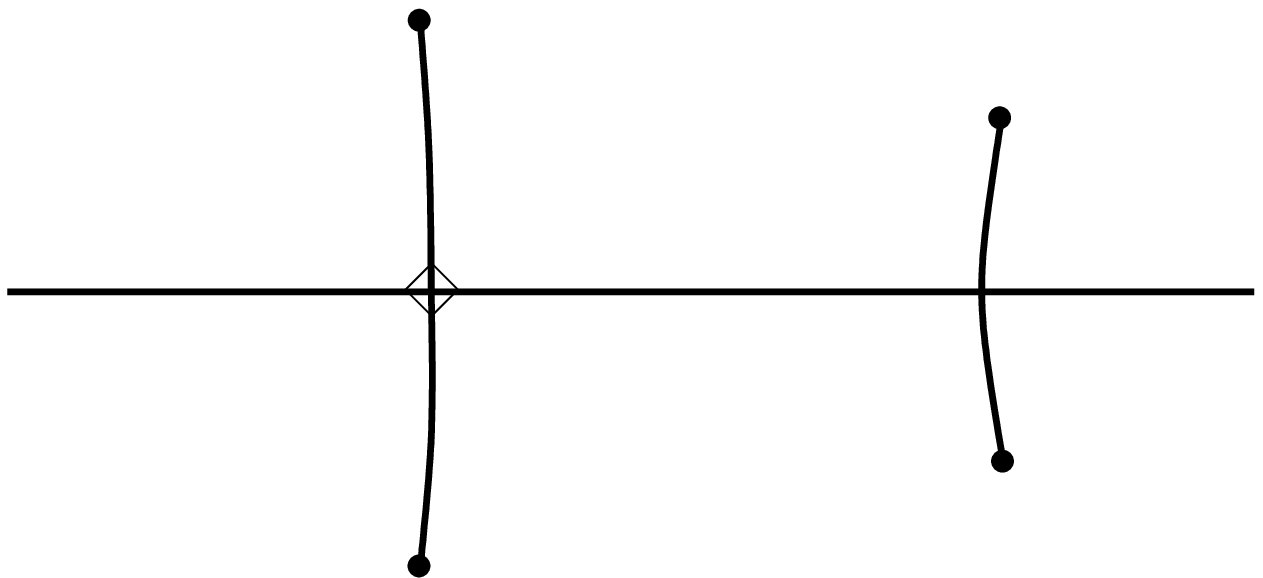}
\vspace{1in}\end{minipage}

\bigskip
\includegraphics[width=3.5in]{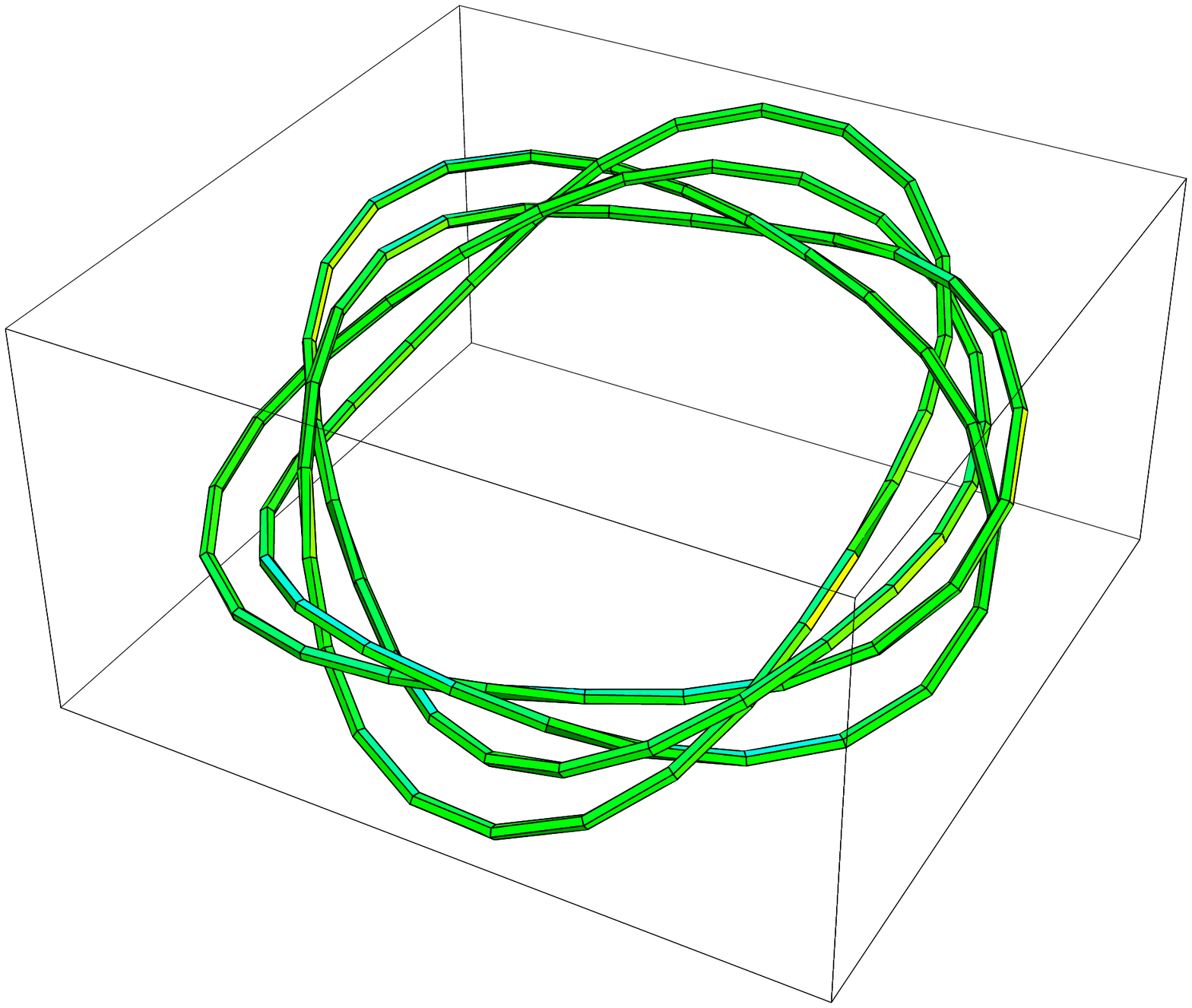}
\quad\begin{minipage}[b]{2.5in}\vspace{.5in}
\includegraphics[width=2.5in]{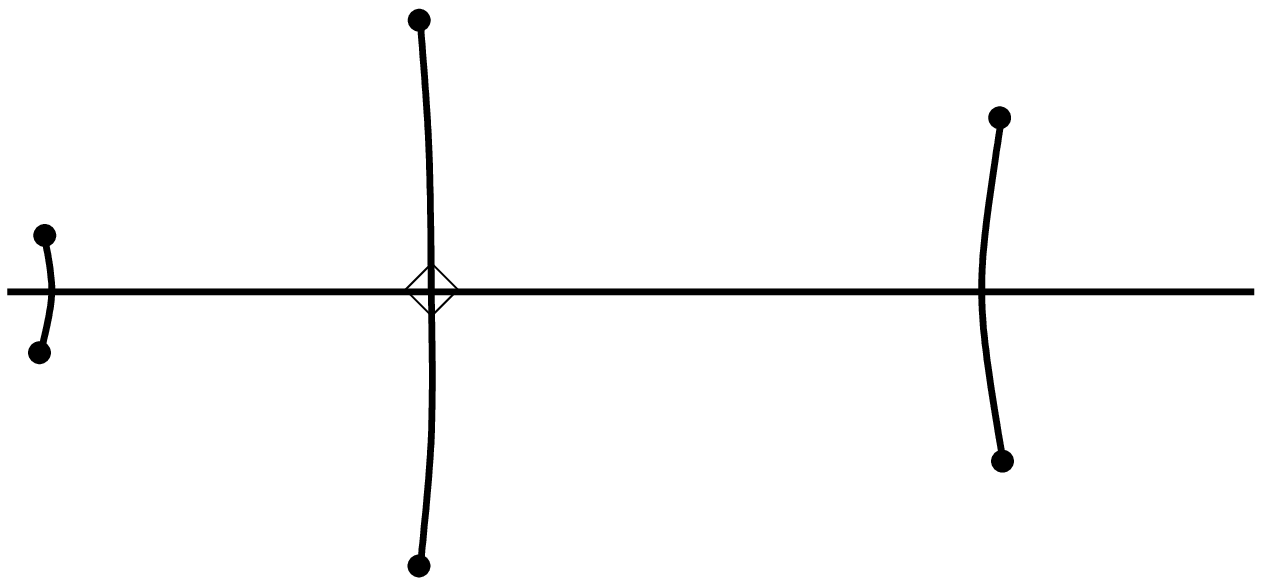}
\vspace{1in}\end{minipage}

\caption{Results of the two-step deformation $[4;-10,7]$.  The end of the first step
is the left-hand $(2,-5)$ torus knot shown at top, and the result of the second
step is the knot shown at bottom, which is a right-hand $(2,7)$ cable on the torus knot.
Approximate diagrams of the Floquet spectrum appear at right; in these, the reconstruction
point is marked by a diamond, while double points are not shown.}\label{yellowgreen}
\end{figure}

\begin{figure}
\centering
\includegraphics[width=3in]{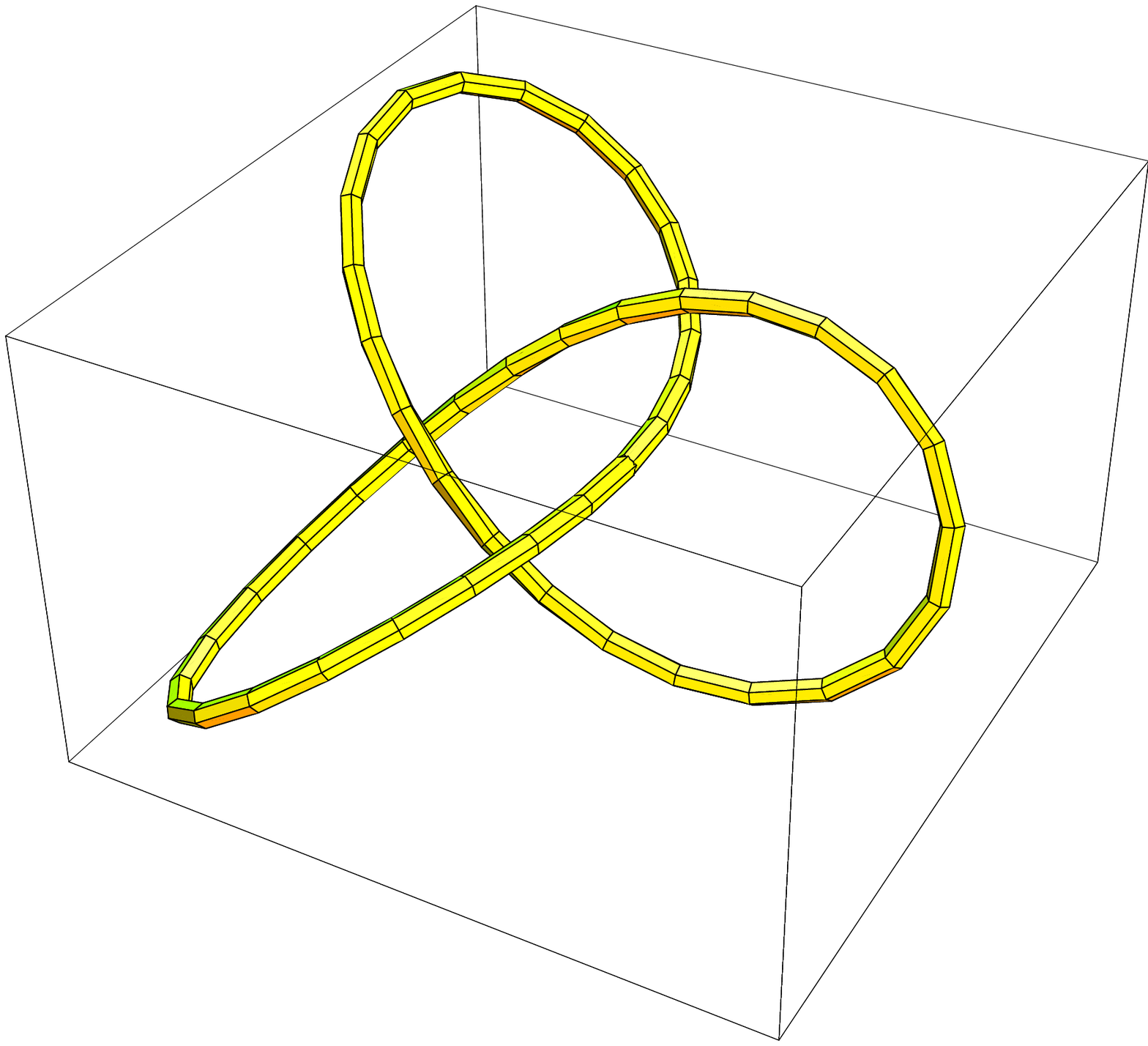}
\quad\begin{minipage}[b]{2in}\vspace{.5in}
\includegraphics[width=2in]{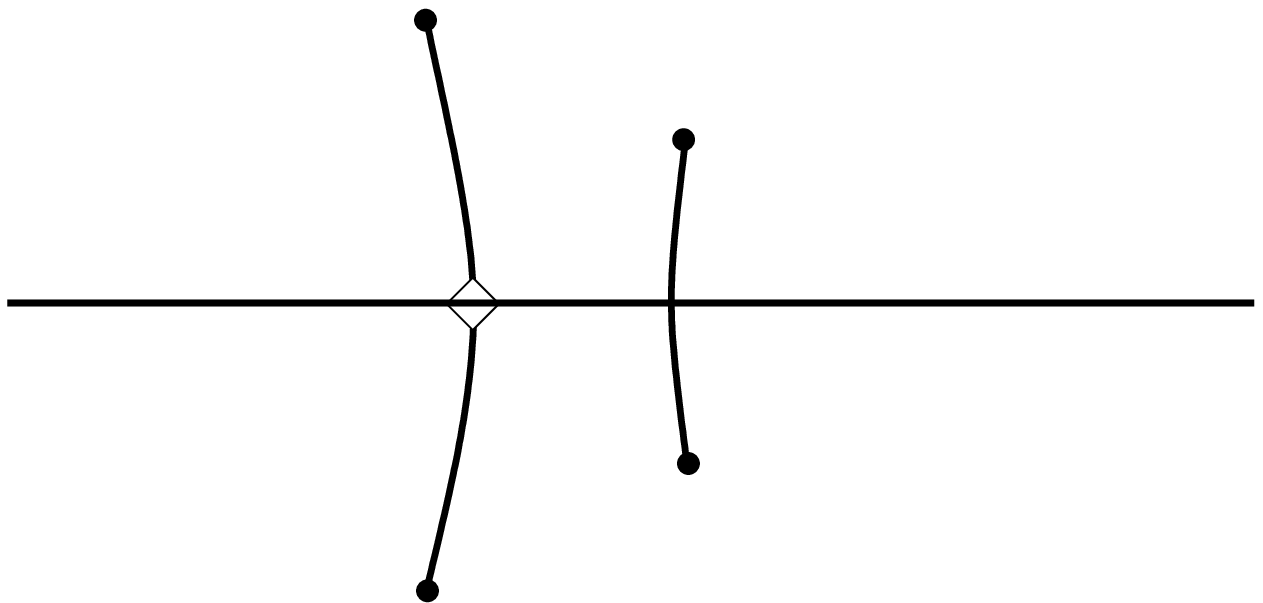}
\vspace{1in}\end{minipage}

\vskip -.3in
\includegraphics[width=3in]{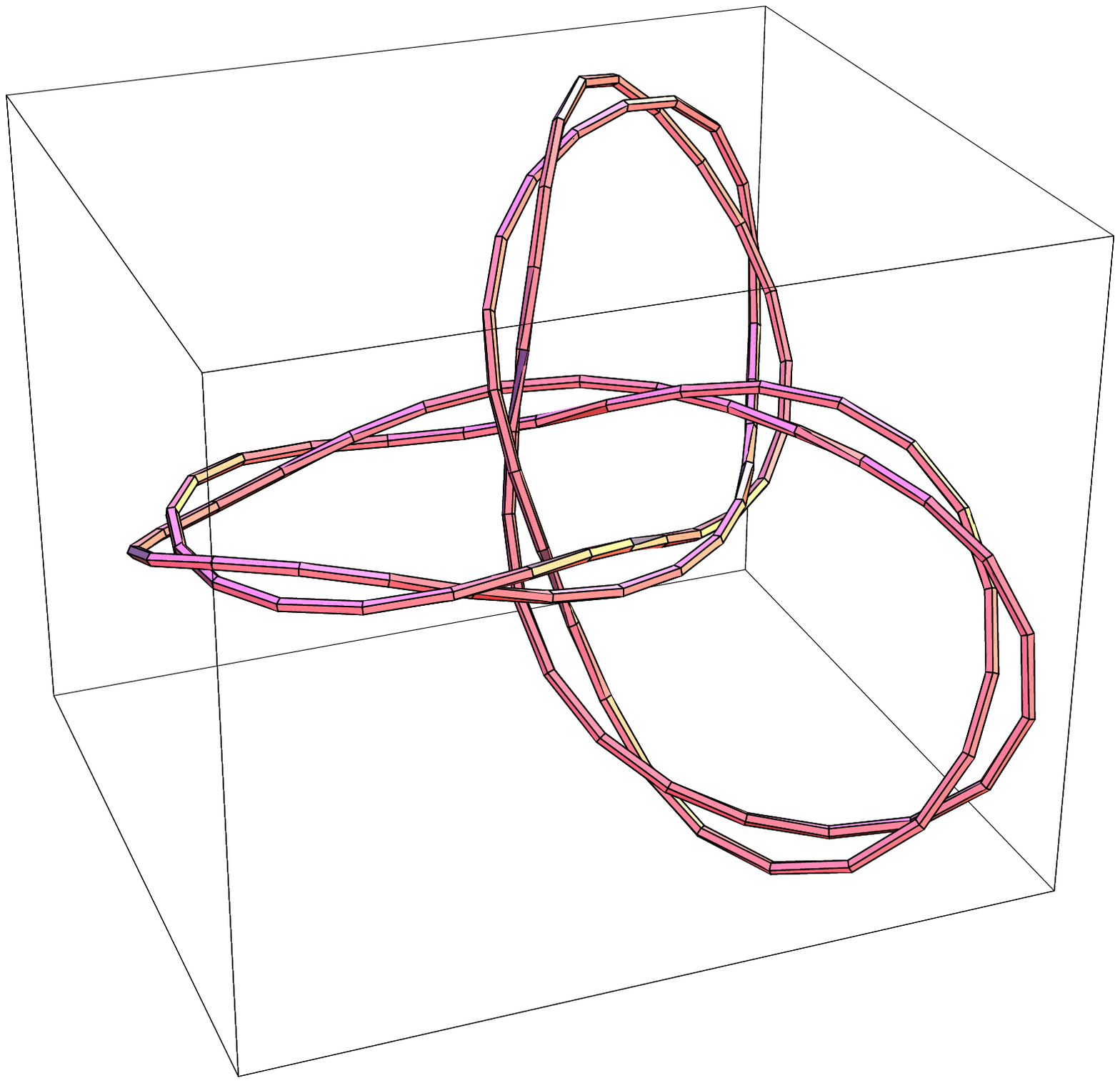}%
\quad\begin{minipage}[b]{2.5in}\vspace{.5in}
\includegraphics[width=2.5in]{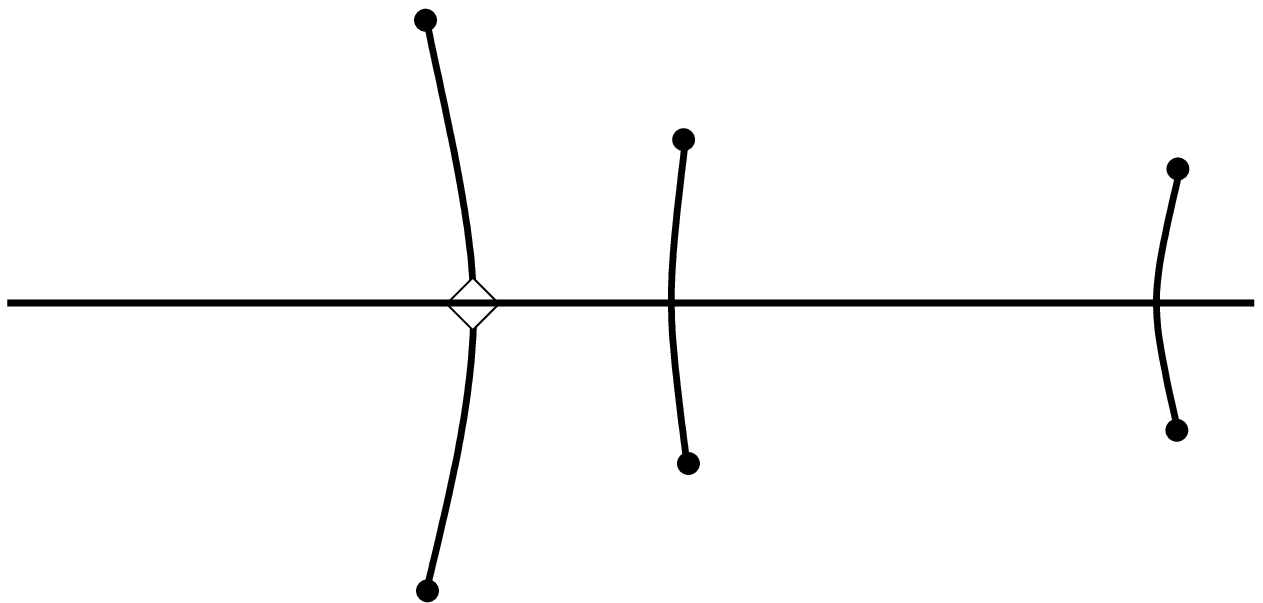}
\vspace{1in}\end{minipage}

\vskip -.3in
\includegraphics[width=3in]{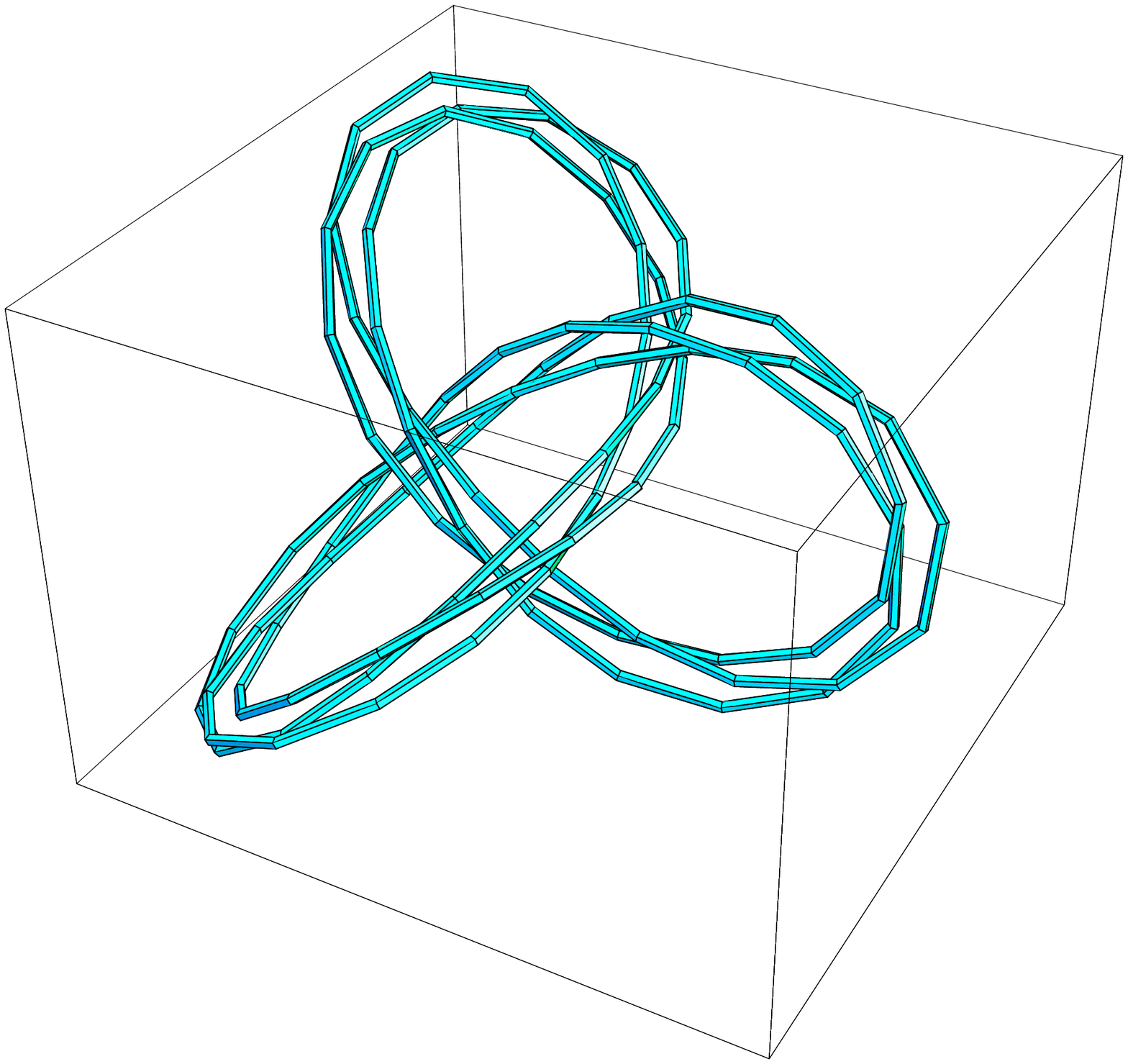}
\quad\begin{minipage}[b]{2.5in}\vspace{.5in}
\includegraphics[width=2.5in]{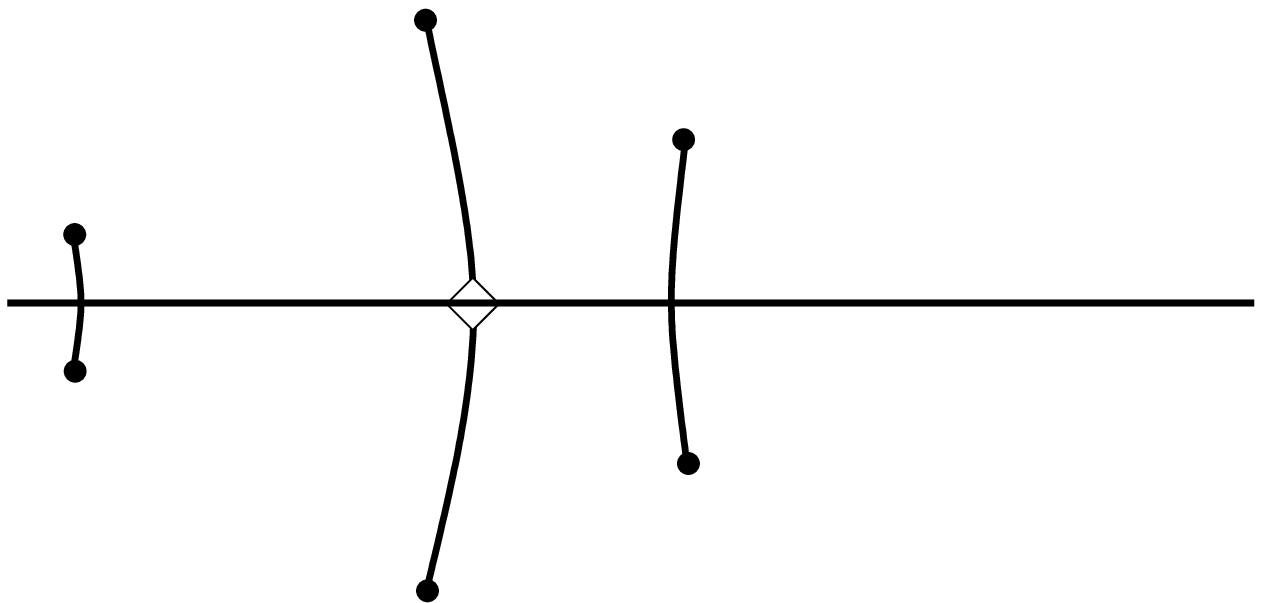}
\vspace{1in}\end{minipage}

\caption{Results of the two-step deformations $[4;-6,-13]$ (in the middle) and $[6;-9,10]$ (at bottom).
In both cases, the result of the first step of the deformation gives the left-hand trefoil shown at top.
The $[4;-6,-13]$ is a left-hand $(2,-13)$ cable on the trefoil, while the $[6;-9,10]$ is a right-hand $(3,5)$
cable on the trefoil.}\label{trefoildefs}
\end{figure}

\begin{figure}
\centering
\includegraphics[width=3in]{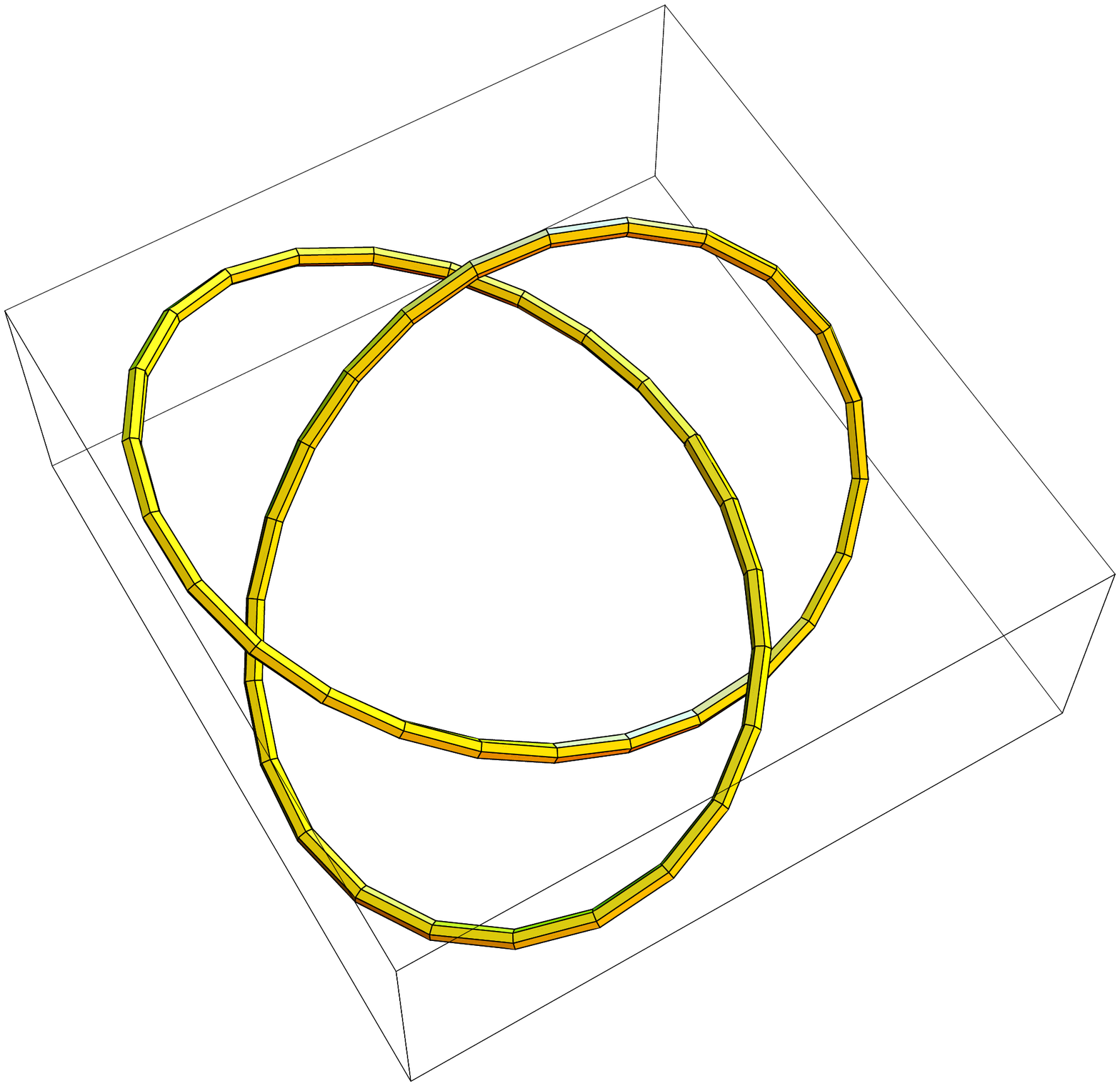}
\begin{minipage}[b]{2.5in}
\includegraphics[width=2in]{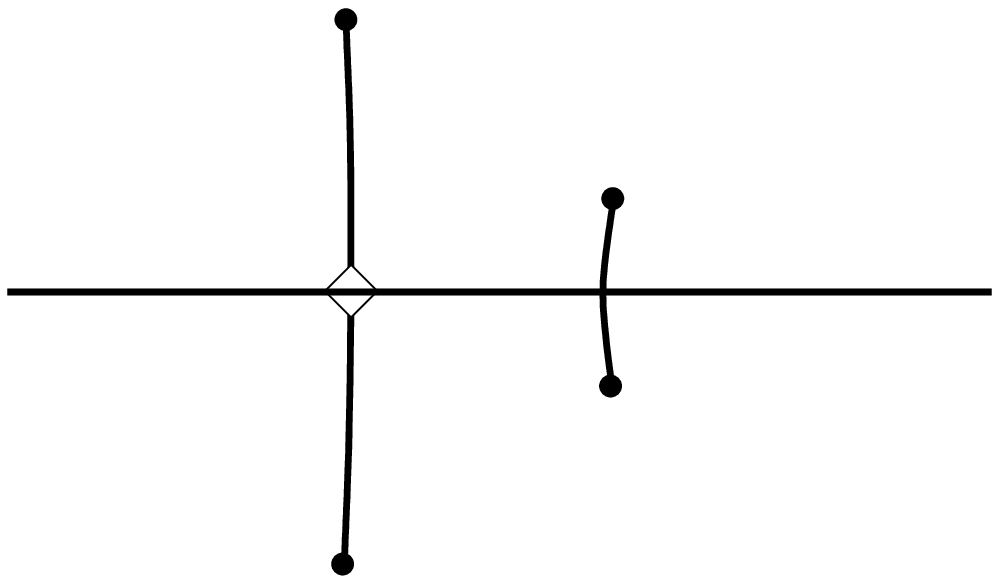}
\vspace{1.5in}\end{minipage}

\vskip -.5in
\includegraphics[width=3in]{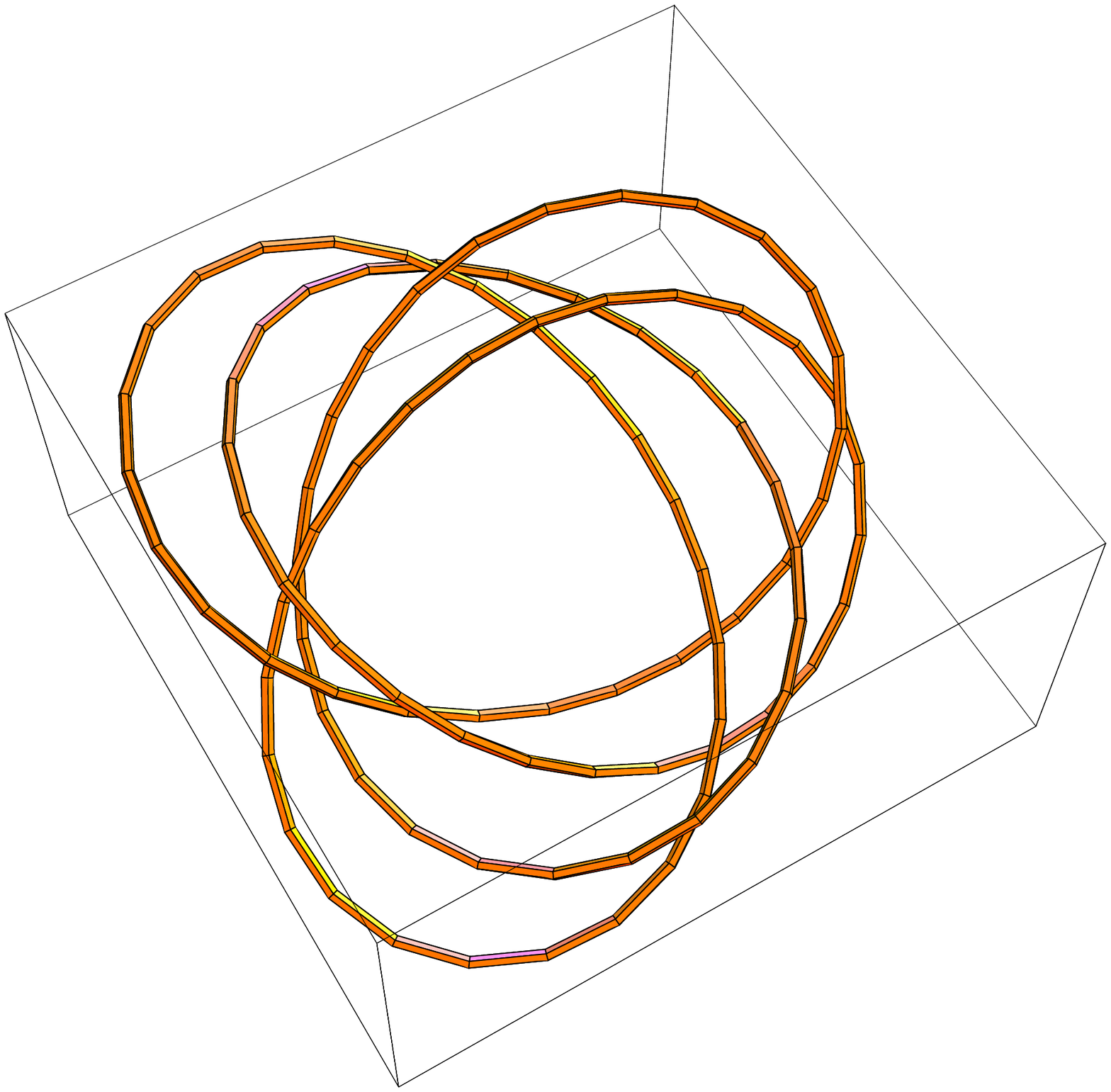}
\begin{minipage}[b]{2.5in}
\includegraphics[width=2in]{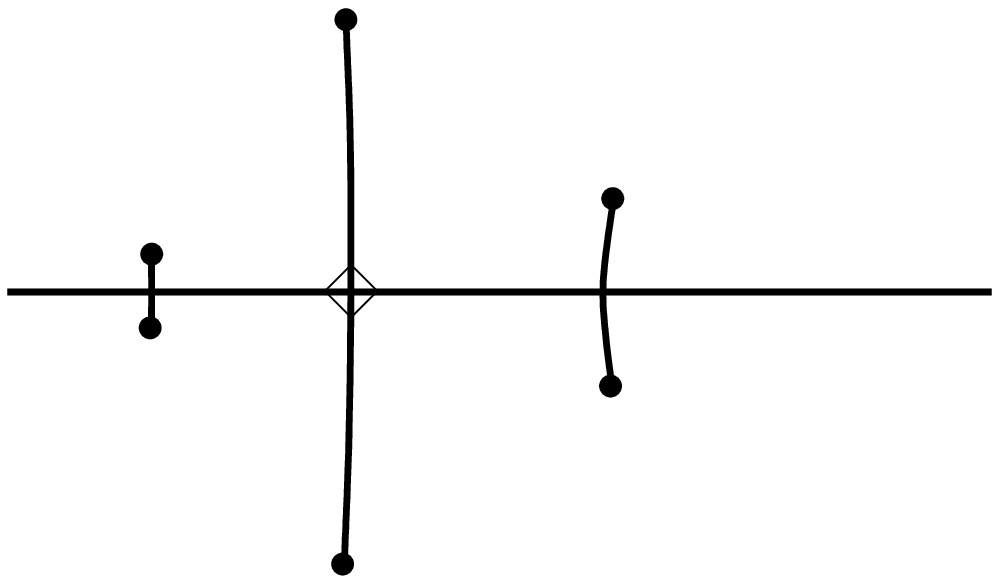}
\vspace{1.5in}\end{minipage}

\vskip -.6in
\includegraphics[width=4in]{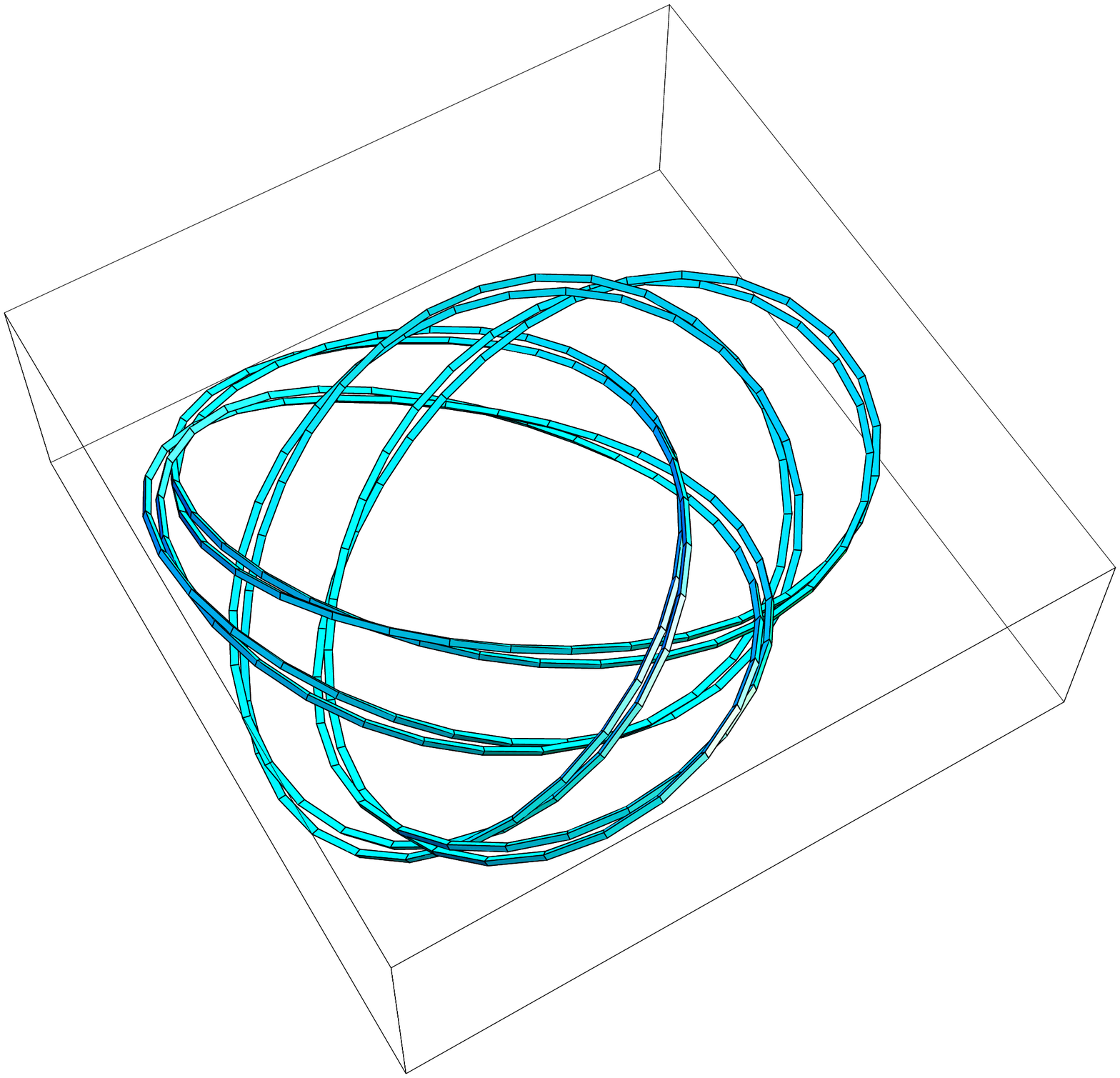}
\hskip -.9in\begin{minipage}[b]{2.5in}
\includegraphics[width=2in]{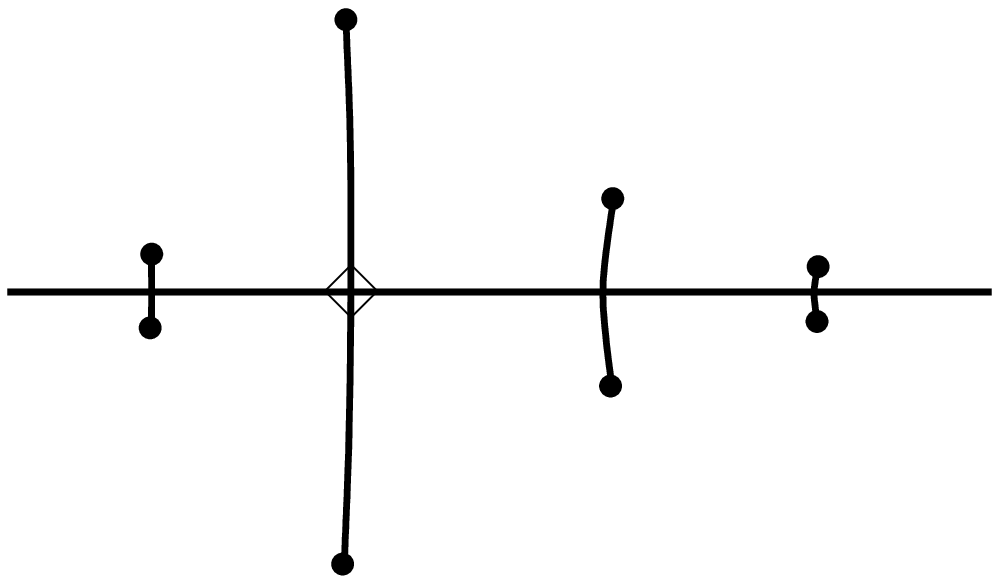}
\vspace{2.5in}\end{minipage}

\vskip -.3in
\caption{Results of the three-step deformation $[8;-12,10,-17]$.  The first step gives
the left-hand trefoil shown at top, the second step (shown in the middle) gives a right-hand $(2,5)$ cable
on the trefoil, and the third step (shown at bottom) gives a left-hand $(2,-17)$ cable on the previous cable.}\label{threestep}
\end{figure}

As we will eventually show, the selection of frequencies determines the knot type
of the resulting filament as an iterated cable knot.  This can be viewed as a generalization
of the work of Keener \cite{Ke}, who showed that if one adds to a circle of length
$L$ a small perturbation of period $(n/m)L$, and a closed curve results, then
the perturbed circle is a $(n,m)$ torus knot (i.e., it covers the original
circle $n$ times lengthwise, and wraps around the circle $m$ times).
The Figures \ref{yellowgreen} through \ref{threestep} show examples of our iterated cable construction.

%% file: isoperturb.tex
 \section{Isoperiodic Deformations and Squared Eigenfunctions}
 \label{isoperturb}

In this section we show how, for an NLS solution $q_0$ of a given period,
the only isoperiodic deformations that preserve the reality condition
$r=-\bar{q}$ (see \S\ref{isoapp1} for notation) and open up just one real
double point $\lambda_0$, while leaving all other points of the discrete
spectrum and the critical points unchanged at first order, must be linear
combinations with real coefficients of
$(\varphi^+)_1^2+(\bar{\varphi_2}^+)^2$ and $\ri[(\varphi^+)_1^2-(\bar{\varphi_2}^+)^2]$,
written in terms of the
components of a Bloch eigenfunction $\vbloch^+$ evaluated at
$\lambda_0$.
The proof requires computing the first order variations of  the discrete
spectrum (simple and multiple points) and of the critical points. For the
purpose of this paper, it is sufficient to discuss the computation of the
first order variation of a real double point; the cases of simple and
critical points can be treated analogously.

Assume that $(q_0+\epsilon q_1, r_0+\epsilon r_1)$ is a periodic
perturbation of a NLS potential $\mathbf{q}_0=(q_0, r_0)$ of a fixed
period $L$, and assume that the pair $(\vphi_0, \lambda_0)$ solves the
AKNS eigenvalue problem at $\mathbf{q}_0$,  where $\lambda_0$ is a real
double point in the spectrum of the unperturbed potential $\mathbf{q}_0$.
Let $\vphi=\vphi_0+\epsilon \vphi_1$ and
$\lambda=\lambda_0+\epsilon\lambda_1$ be the corresponding variations of eigenfunction and
eigenvalue. At first order in $\epsilon$, we obtain
\begin{equation}
\label{firstorderLS}
\mathcal{L}_1\vphi_1=\begin{pmatrix} -\ri \lambda_1 & \ri q_1 \\ -\ri r_1 & \ri \lambda_1    \end{pmatrix}
\vphi_0,
\end{equation}
where $\displaystyle \mathcal{L}_1 = \frac{\rd}{\rd x} - \begin{pmatrix} -\ri \lambda_0 & \ri q_0 \\ -\ri r_0 & \ri \lambda_0    \end{pmatrix}$
is the spatial operator of the unperturbed AKNS linear system at $\lambda_0$.

Together with the homogeneous linear system $\mathcal{L}_1\vphi_0=\mathbf{0}$, we consider its formal adjoint with respect to the $L^2$-inner product
$\displaystyle \left\langle \mathbf{u}, \mathbf{v}\right\rangle=\int_0^L \mathbf{u} \cdot \overline{\mathbf{v} }\, \rd x$:
\[
\mathcal{L}^H_1\vpsi=\mathbf{0}, \qquad \qquad \mathrm{with} \qquad \qquad  \mathcal{L}^H_1=-\frac{\rd}{\rd x} + \begin{pmatrix} -\ri \bar{\lambda}_0 & -\ri \bar{r}_0 \\ \ri \bar{q}_0 & \ri \bar{\lambda}_0 \end{pmatrix}.
\]
(The ``formal adjoint" is computed neglecting boundary conditions. It turns out that the solvability condition of system \eqref{firstorderLS}
 is correctly expressed in terms of the formal adjoint, as discussed below.)
Taking the inner product of both sides of \eqref{firstorderLS} with a solution of $\mathcal{L}^H_1\vpsi=\mathbf{0}$ leads to  the
{\sl solvability condition} for system \eqref{firstorderLS}, requiring that its right-hand side is orthogonal
to any solution of the  homogeneous formal adjoint system.

Since $\lambda_0$ is  assumed to be a real double point, it is removable \cite{Schm}.
This means that the space of solutions of $\mathcal{L}_1\vphi=\mathbf{0}$ is spanned by the two linearly independent Bloch eigenfunctions
(see \S\ref{isoapp1})
\[
\boldsymbol{\varphi}^+=\left( \begin{array}{c} \varphi_1^+ \\ \varphi_2^+ \end{array} \right),
    \qquad \boldsymbol{\varphi}^-=\left( \begin{array}{c} \varphi_1^- \\ \varphi_2^- \end{array} \right),
\]
evaluated at $\lambda_0$.
It is easy to show that a basis for the solution space of the adjoint system $\mathcal{L}_1^H\vpsi=\mathbf{0}$ is given by
\[
J\overline{\boldsymbol{\varphi}^+}=\left( \begin{array}{c} \overline{\varphi_2^+} \\ -\overline{\varphi_1^+} \end{array} \right),
   \qquad J\overline{\boldsymbol{\varphi}^-}=\left( \begin{array}{c} \overline{\varphi_2^-} \\ -\overline{\varphi_1^-} \end{array} \right),
\]
where $J$ is the symplectic matrix $\begin{pmatrix} 0 & -1 \\ 1 & 0 \end{pmatrix}$.
Substituting  ${\vphi}_0=a\mathbf{\varphi}^++b \mathbf{\varphi}^-$, a general solution of $\mathcal{L}_1\vphi=\mathbf{0}$,
in system \eqref{firstorderLS},
we compute the solvability condition to be the following system in the unknowns $a$ and $b$:
\[
\footnotesize
\begin{split}
\left[  2\ri \lambda_1 \int_0^L \varphi_1^+\varphi_2^+ \, \rd x + \ri \left\langle \mathbf{q}_1, J \begin{pmatrix} (\overline{\varphi_1^+})^2 \\ - (\overline{\varphi_2^+})^2   \end{pmatrix}\right\rangle \right] a +\left[  \ri \lambda_1 \int_0^L (\varphi_1^+\varphi_2^-+\varphi_2^+\varphi_1^-) \, \rd x + \ri \left\langle \mathbf{q}_1, J \begin{pmatrix} \overline{\varphi_1^+} \overline{\varphi_1^-} \\ - \overline{\varphi_2^+}   \overline{\varphi_2^-}\end{pmatrix} \right\rangle \right] b =0, \\
\left[ \ri \lambda_1 \int_0^L (\varphi_1^+\varphi_2^-+\varphi_2^+\varphi_1^-) \, \rd x + \ri \left\langle \mathbf{q}_1, J  \begin{pmatrix} \overline{\varphi_1^+} \overline{\varphi_1^-} \\ - \overline{\varphi_2^+}   \overline{\varphi_2^-}\end{pmatrix} \right\rangle \right] a+\left[ 2\ri \lambda_1 \int_0^L \varphi_1^-\varphi_2^- \, \rd x + \ri \left\langle \mathbf{q}_1, J   \begin{pmatrix} (\overline{\varphi_1^-})^2 \\ - (\overline{\varphi_2^-})^2   \end{pmatrix}\right\rangle
 \right] b =0,
\end{split}
\]
where the various components of the Bloch eigenfunctions are evaluated at the double points $\lambda_0$.

For a removable double point $\lambda_d$, the following identities hold
\[
\int_0^L \left. \varphi_1^+\varphi_2^+ \right|_{\lambda_d} \, \rd x=0, \qquad \qquad \int_0^L \left. \varphi_1^-\varphi_2^- \right|_{\lambda_d}\, \rd x=0,
\]
as shown, for example, in \cite{GSc}. Thus, the condition for  existence of a nontrivial solution $(a,b)$
of the above system (amounting to the vanishing of the determinant of the associated matrix) provides the selection of
the perturbation of the affected double point. A simple calculation gives
\[
\small
\lambda_1=\frac{\left\langle \mathbf{q}_1, J \left. \begin{pmatrix} \overline{\varphi_1^+} \overline{\varphi_1^-} \\ - \overline{\varphi_2^+}   \overline{\varphi_2^-} \end{pmatrix} \right|_{\lambda_0}\right\rangle \pm
\sqrt{\left\langle \mathbf{q}_1, J \left. \begin{pmatrix} (\overline{\varphi_1^+})^2 \\ - (\overline{\varphi_2^+})^2   \end{pmatrix}\right|_{\lambda_0} \right\rangle \left\langle \mathbf{q}_1, J   \left. \begin{pmatrix} (\overline{\varphi_1^-})^2 \\ - (\overline{\varphi_2^-})^2   \end{pmatrix}\right|_{\lambda_0} \right\rangle}}{\int_0^L \left. (\varphi_1^+\varphi_2^-+\varphi_2^+\varphi_1^-)  \right|_{\lambda_0}\, \rd x},
\]
where $\mathbf{q}_1$ denotes the vector $(q_1,r_1)^T.$

We now recall that certain vectors of quadratic products of components of the Bloch eigenfunctions (the {\sl squared eigenfunctions})
form a basis for $L^2_{per}([0,L],\mathbb{C})$. (See \S\ref{isoapp1} and references therein.) Moreover, we can write a
generic periodic perturbation $\mathbf{q}_1$ as a linear combination of the elements of the basis of {\sl squared eigenfunctions}:
\[
\small
\begin{split}
\begin{pmatrix} q_1 \\ r_1 \end{pmatrix}= \sum_{\text{simple \, points\,}  \lambda_s} c_s \left. \begin{pmatrix} \phi_1^2 \\ -\phi_2^2 \end{pmatrix} \right|_{\lambda_s} + &   \sum_{\text{critical \, non-per.\, pts \,}  \lambda_c} d_s \left. \begin{pmatrix} \varphi^+_1 \varphi^-_1\\ -\varphi^+_2\varphi^-_2 \end{pmatrix} \right|_{\lambda_c} \\ +  &\sum_{\text{remov. \, double \, pts. \,} \lambda_d}  \left. [e^+_d  \begin{pmatrix} (\varphi^+_1)^2 \\ -(\varphi^+_2)^2 \end{pmatrix} +e^-_d  \begin{pmatrix} (\varphi^-_1)^2 \\ -(\varphi^-_2)^2 \end{pmatrix} ] \right|_{\lambda_d}.
\end{split}
\]
(We have assumed for simplicity that the potential $\mathbf{q}_0$ possesses no non-removable double points and no periodic points of multiplicity higher than two.)

Because of the biorthogonality property of the squared eigenfunctions (see \S\ref{isoapp1}), we compute
\[
\left\langle \mathbf{q}_1,  J \left. \begin{pmatrix} \overline{\varphi_1^+ \varphi_1^-}\\ - \overline{\varphi_2^+ \varphi_2^-}\   \end{pmatrix} \right|_{\lambda_0} \right\rangle =0,
\]
\[\footnotesize
\left\langle \mathbf{q}_1, J \left. \begin{pmatrix} (\overline{\varphi_1^+})^2 \\ - (\overline{\varphi_2^+})^2   \end{pmatrix} \right|_{\lambda_0} \right\rangle =e_0^- \left. \left\langle \begin{pmatrix} (\varphi_1^-)^2 \\ - (\varphi_2^-)^2   \end{pmatrix}, J \begin{pmatrix} (\overline{\varphi_1^+})^2 \\ - (\overline{\varphi_2^+})^2   \end{pmatrix}\right\rangle \right|_{\lambda_0} =-e_0^-  \left. \overline{
\left\langle \begin{pmatrix} (\varphi_1^+)^2 \\ - (\varphi_2^+)^2   \end{pmatrix}, J \begin{pmatrix} (\overline{\varphi_1^-})^2 \\ - (\overline{\varphi_2^-})^2   \end{pmatrix}\right\rangle} \right|_{\lambda_0}
\]
and
\[
 \left\langle \mathbf{q}_1, J  \left. \begin{pmatrix} (\overline{\varphi_1^-})^2 \\ - (\overline{\varphi_2^-})^2   \end{pmatrix} \right|_{\lambda_0} \right\rangle=e^+_0 \left. \left\langle \begin{pmatrix} (\varphi_1^+)^2 \\ - (\varphi_2^+)^2   \end{pmatrix}, J \begin{pmatrix} (\overline{\varphi_1^-})^2 \\ - (\overline{\varphi_2^-})^2   \end{pmatrix}\right\rangle \right|_{\lambda_0},
\]
with normalization coefficient (see Lemma \ref{biopair})
\[
\left. \left\langle \begin{pmatrix} (\varphi_1^+)^2 \\ - (\varphi_2^+)^2   \end{pmatrix}, J \begin{pmatrix} (\overline{\varphi_1^-})^2 \\ - (\overline{\varphi_2^-})^2   \end{pmatrix} \right\rangle \right|_{\lambda_0}
=\frac{1}{2\ri} \sqrt{\Delta (\lambda_0) \Delta''(\lambda_0 )}\left| W[\boldsymbol{\varphi}^+,\boldsymbol{\varphi}^-] \right|^2_{\lambda_0},
\]
where $W[\boldsymbol{\varphi}^+,\boldsymbol{\varphi}^-]$ denotes the Wronskian of the two Bloch eigenfunctions.
Because the normalization coefficient is non zero, it follows that the real double point $\lambda_0$ will split
at first order and none of the remaining double point will if and only if the perturbation
$\mathbf{q}_1$ contains only terms of the form
$$ \left. e^+_0  \begin{pmatrix} (\varphi^+_1)^2 \\ -(\varphi^+_2)^2 \end{pmatrix} \right|_{\lambda_0} +e^-_0  \left. \begin{pmatrix} (\varphi^-_1)^2 \\ -(\varphi^-_2)^2 \end{pmatrix}        \right|_{\lambda_0}.
$$
and  none of  the terms
$ \left. e^+_d  \begin{pmatrix} (\varphi^+_1)^2 \\ -(\varphi^+_2)^2 \end{pmatrix} \right|_{\lambda_d} +e^-_d  \left. \begin{pmatrix} (\varphi^-_1)^2 \\ -(\varphi^-_2)^2 \end{pmatrix}        \right|_{\lambda_d}$.

Analogous results can be deduced by computing expressions for the first order variation  of simple and critical points:
   they will move at first order  if and only if the potential $\mathbf{q}_1$ contains terms of the form
   $\displaystyle \left. \begin{pmatrix} \varphi_1^2 \\ -\varphi_2^2 \end{pmatrix} \right|_{\lambda_s}$ or
   $\displaystyle  \left. \begin{pmatrix} \varphi^+_1 \varphi^-_1\\ -\varphi^+_2\varphi^-_2 \end{pmatrix} \right|_{\lambda_c} $
   respectively. Thus, because we are assuming that these points do not move to first order, then these terms do not occur in $\mathbf{q}_1$.

Since $\lambda_0$ is assumed to be real, the Bloch eigenfunctions possess the additional symmetry $\boldsymbol{\varphi}^- =J \overline{\boldsymbol{\varphi}^+}$, thus a perturbation $\mathbf{q}_1$ that splits only $\lambda_0$, leaving the rest of the discrete spectrum and critical points invariant at first order must be of the form:
\[
\begin{pmatrix} q_1 \\ r_1 \end{pmatrix} = \left. \begin{pmatrix} e^+_0 (\varphi^+_1)^2 +e^-_0 (\varphi^+_2)^2 \\
-[e^+_0 (\varphi^+_2)^2 +e^-_0 (\varphi^+_1)^2] \end{pmatrix} \right|_{\lambda_0}.
\]
Finally, requiring that the focussing reality constraints $r_1=-\bar{q}_1$ be satisfied leads to the condition
\[
(e^+_0-\overline{e^-_0}) (\varphi^+_2)^2-(\overline{e^+_0}-e^-_0) (\varphi^+_1)^2=0,
\]
and, given the linear independence of $(\varphi^+_1)^2$ and $(\varphi^+_2)^2$ over $\mathbb{C}$,
one obtains $e^+_0=\overline{e^-_0}=c$ for some complex constant $c$. We summarize these results  in the following
\begin{proposition}
\label{formofq1}
Suppose the periodic potential $q_0$ undergoes a smooth perturbation $q=q_0+\epsilon q_1$
and its spectrum deforms isoperiodically in such a way a unique real double point $\lambda_0$ splits at first order,
while all other points the discrete spectrum as well as the critical points remain unchanged up to first order.
If, in addition, the deformation preserves the reality of the potential, then the perturbation must have the form
$q_1=\left.\realpart{(c)}[(\varphi_1^+)^2+(\bar{\varphi}_2^+)^2]\right|_{\lambda_0}
+\left.\ri \impart{(c)}[(\varphi_1^+)^2-(\bar{\varphi_2^+})^2]\right|_{\lambda_0}$, with $c \in \mathbb{C}$.
Moreover, the affected double point splits at first order in the following pair of simple points
\[
\lambda_\pm=\lambda_0 \pm \ri \epsilon |c|^2  \frac{|W[\boldsymbol{\varphi}^+,\boldsymbol{\varphi}^-]|^2 \sqrt{|\Delta''(\lambda_0)|}}{\int_0^L (|\varphi_1^+|^2-|\varphi_2^+|^2) \, \rd x}  + O(\epsilon^2).
\]
\end{proposition}

%% file: isocabling.tex
\section{The Cabling Theorem}\label{isocabling}
%
%
In this section we will assume that $q_0$ is an NLS potential of period $n\pi$, obtained by
a sequence of $g$ homotopic deformations from that of the $n$-times covered circle, notated
as $[n;m_1,\ldots, m_g]$, and that $\lambda_0$ is a double point of the $j n\pi$-periodic
spectrum of $q_0$ whose original position (in the spectrum of the circle, before the deformations)
was $-\sqrt{(m/(jn))^2 -1}$ times the sign of $m$,  where $m$ is an integer whose magnitude is greater that $j$ and
which is relatively prime to $j$.  We will show that, if we perturb $q_0$ as in Proposition
\ref{formofq1}, then the perturbed curve $\vgamma=\vgamma_0+\epsilon \vgamma_1 + O(\epsilon^2)$ is an $(j,m)$-cable
on $\vgamma_0$ for sufficiently small $\epsilon$.

We also need to assume that $q_0$ is sufficiently close to the plane wave potential; we will
be more specific about this assumption at the end of this section.

\subsection{Perturbed Potentials and Perturbed Curves via the Sym Formula}


We will begin by calculating the perturbation of the curve.
We assume that
$$q(x,t;\epsilon) = q_0 + \epsilon q_1 + O(\epsilon^2)$$
is a one-parameter family of NLS solutions,
and
$$\Rphi(x,t;\lambda,\epsilon) = \Rphi_0 + \epsilon \Rphi_1 + O(\epsilon^2),$$
is a fundamental matrix solution for the AKNS system at $(q,\lambda)$.  To be specific,
we will assume that $\Rphi(0,0;\lambda,\epsilon)$ is the identity matrix.
Moreover, when $\lambda$ is real, we can assume that $\Rphi$ takes
the form $[\vrphi, J \widebar\vrphi]$.
We will also assume that $q_1$ is expressed as a linear combination of squared eigenfunctions
for the AKNS system at $(q_0,\lambda_0)$, where $\lambda_0$ is a double point for $q_0$.
(This assumption was justified, for isoperiodic deformations
that open up a double point, in the previous section.)

For use in the Sym-Pohlmeyer formula \eqref{RECO}, we will need to calculate $\Rphi_1$ at $\lambda=\Lambda_0$,
the reconstruction point.  (Although this reconstruction point will vary under isoperiodic deformation,
Proposition \ref{hardprop} ensures that it is fixed up to $O(\epsilon^2)$.)
For the sake of brevity, we will use the abbreviations
$$
\bloch_1 = \bloch^+_1(q_0,\lambda_0), \quad \bloch_2 = \bloch^+_2(q_0,\lambda_0),
$$
where $\vbloch^+$ is one of the Bloch eigenfunctions,
as normalized in
\eqref{normalizedbloch}.
Then the results of \S\ref{isoperturb} imply that
\begin{equation}\label{q1form}
q_1 = c(\bloch_1)^2 + \overline{c}(\overline\bloch_2)^2
\end{equation}
for some complex constant $c$.  We will also let
\begin{equation}\label{Phiform}
\Rphi_0(x,t;\lambda) = \begin{pmatrix} \rphi_1 &  -\widebar\rphi_2 \\ \rphi_2 & \widebar\rphi_1 \end{pmatrix},
\end{equation}
where we will leave $\lambda$ arbitrary (but real) for the moment, but later set $\lambda=\Lambda_0$.

Setting $B_1 = \Rphi_1 (\Rphi_0)^{-1}$ and taking the $\epsilon$ term on each side of the AKNS system gives
$$\dfrac{\rd}{\rd x} B_1 = \Rphi^{-1}_0 Q_1 \Rphi_0, \qquad Q_1 = \begin{pmatrix} 0 & \ri q_1 \\ \ri \widebar q_1 & 0 \end{pmatrix}$$
(see \cite{CI3} for more details).  Expanding the inverse, we get
\begin{equation}\label{bone}
\dfrac{\rd}{\rd x} B_1 = \dfrac{\ri}{D}
\begin{pmatrix}q_1 \widebar\rphi_1 \rphi_2 + \widebar q_1 \rphi_1 \widebar\rphi_2 & q_1 \widebar\rphi_1^2 -\widebar q_1\widebar\rphi_2^2 \\
\widebar q_1 \rphi_1^2 -q_1 \rphi_2^2 &  -q_1 \widebar\rphi_1 \rphi_2 -\widebar q_1 \rphi_1 \widebar\rphi_2
\end{pmatrix},
\end{equation}
where $D = \det \Rphi_0 = |\rphi_1|^2+|\rphi_2|^2$, which is independent of $x$ and $t$.

To compute $B_1$, we need to take antiderivatives of the
entries of the matrix appearing on the right-hand side of \eqref{bone}.
(Because the matrix takes value in $\mathfrak{su}(2)$, it suffices to
find antiderivatives for entries in the first column.)
To do this we will use special properties of solutions of the squared
eigenfunction system \eqref{fghxsys},\eqref{fghtsys}
Recall  that if $\vrphi,\widetilde\vrphi$ are
two vector solutions of the AKNS system at the same $\lambda$-value, then the construction
\begin{equation}\label{triplese}
\begin{bmatrix} f \\ g\\ h \end{bmatrix}
= \vrphi \otimes \widetilde\vrphi = \begin{bmatrix} \tfrac12 (\phi_1 \widetilde\rphi_2 + \widetilde\rphi_1 \rphi_2) \\
\rphi_1 \widetilde \rphi_1 \\ -\rphi_2 \widetilde\rphi_2 \end{bmatrix}
\end{equation}
gives a solution of the squared eigenfunction system \eqref{fghxsys},\eqref{fghtsys}
(see \S\ref{squaredapp} in the Appendix).
Furthermore, if $(f,g,h)$ and $(\widehat f, \widehat g, \widehat h)$ are solutions of this system
at $(q_0,\lambda)$ and $(q_0,\widehat \lambda)$ respectively, then
\begin{equation}\label{anna}
\dfrac{\rd}{\rd x} (g \widehat h + \widehat g h -2 f \widehat f)= 2\ri (\lambda- \widehat\lambda) (\widehat g h - g \widehat h).
\end{equation}
(This can be easily verified from the system of ODE's \eqref{fghxsys} satisfied by $(f,g,h)$;
see, for example, the proof of Theorem 2.1 in \cite{MOv}.)

To get the top left entry in \eqref{bone}, let $\widehat g=q_1$, $\widehat h=-\widebar q_1$, $\widehat \lambda=\lambda_0$,
and $g= \rphi_1\widebar\rphi_2$, $h = \rphi_2\widebar\rphi_1$ (the latter arising from the construction $\boldsymbol{\rphi} \otimes (J \overline{\boldsymbol{\rphi}})$,
giving $f = -\tfrac12 (|\rphi_1|^2 - |\rphi_2|^2)$).
Then
$$\int q_1 \widebar\rphi_1 \rphi_2 + \widebar q_1 \rphi_1 \widebar\rphi_2\  \rd x =
\dfrac{1}{2\ri (\lambda - \lambda_0) }
\left(q_1 \widebar\rphi_1 \rphi_2 -\widebar q_1 \rphi_1\widebar\rphi_2
-(|\rphi_1|^2 - |\rphi_2|^2) \widehat f\right),$$
where, using the formula \eqref{q1form} for $q_1$, we have
$$\widehat f = c \bloch_1 \bloch_2 - \overline{c}\, \overline\bloch_1 \overline\bloch_2.$$

To get the bottom left entry in \eqref{bone} (up to a factor of minus one),
we keep $\widehat f, \widehat g, \widehat h$ the same, and change to $g= \rphi_1^2$ and $h = -\rphi_2^2$ (arising
from the construction $\boldsymbol{\rphi} \otimes \boldsymbol{\rphi}$, giving $f= \rphi_1 \rphi_2$).  Then
$$\int q_1 \rphi_2^2 - \widebar q_1 \rphi_1^2 \ \rd x =
\dfrac{1}{2\ri (\lambda - \lambda_0) }
\left( \widebar q_1 \rphi_1^2 + q_1 \rphi_2^2 -2\rphi_1 \rphi_2 \widehat f\right).$$

So, up to an additive constant, we obtain
$$B_1 = \dfrac1{2(\lambda-\lambda_0)D}
\begin{pmatrix}
-\widebar q_1 \rphi_1\widebar\rphi_2 + q_1 \widebar\rphi_1 \rphi_2
-(|\rphi_1|^2 - |\rphi_2|^2)\widehat f &
q_1 \widebar{\rphi}_1^2 + \widebar{q}_1 \widebar{\rphi}_2^2 +2\widebar{\rphi}_1 \widebar{\rphi}_2 \widehat f
\\
-\widebar q_1 \rphi_1^2 - q_1 \rphi_2^2 +2\rphi_1 \rphi_2 \widehat f
&
\widebar q_1 \rphi_1\widebar\rphi_2 - q_1 \widebar\rphi_1 \rphi_2
+(|\rphi_1|^2 - |\rphi_2|^2)\widehat f
\end{pmatrix}.$$

Substituting $\Rphi = \Rphi_0 (I + \epsilon B_1 + O(\epsilon^2))$ into the Sym-Pohlmeyer formula gives
$$\vgamma = \vgamma_0 + \epsilon \vgamma_1 + O(\epsilon^2), \qquad
\vgamma_0 = \Rphi_0^{-1}\dfrac{\rd\Rphi_0}{\rd\lambda}, \qquad \vgamma_1 =\dfrac{\rd B_1}{\rd\lambda} + [\vgamma_0, B_1].$$
(Recall that we are identifying matrices in $\mathfrak{su}(2)$ with vectors in $\bR^3$.)
But $B_1 =\int \Rphi_0^{-1} Q_1 \Rphi_0\,\rd x$, and $Q_1$ is independent of $\lambda$, so
\begin{align*}
\dfrac{\rd B_1}{\rd\lambda}
&= \int -\vgamma_0 \Rphi_0^{-1} Q_1 \Rphi_0 + \Rphi_0^{-1} Q_1 \dfrac{\rd\Rphi_0}{\rd\lambda}\ \rd x
= \int [ \Rphi_0^{-1} Q_1 \Rphi_0, \vgamma_0]\,\rd x
\\
&= \int [(B_1)_x, \vgamma_0]\, \rd x = [B_1,\vgamma_0] - \int [B_1,T]\, \rd x,
\end{align*}
where integration by parts is used in the second line.  Therefore, $\vgamma_1 = \int [T, B_1]\,\rd x$, where
$$T = \Rphi_0^{-1} \begin{pmatrix} -\ri & 0 \\ \ri & 0 \end{pmatrix} \Rphi_0=
\dfrac{1}{D} \begin{pmatrix} -\ri (|\rphi_1|^2 - |\rphi_2|^2) & 2\ri \widebar\rphi_1 \widebar\rphi_2 \\
2\ri \rphi_1 \rphi_2 & \ri (|\rphi_1|^2 - |\rphi_2|^2) \end{pmatrix}.
$$
We now set $\lambda=\Lambda_0$, the reconstruction point that generates the closed curve $\vgamma_0$ from
potential $q_0$.  Thus, because $\int T \,\rd x = 0$, the additive constant in $B_1$ does not
change the value of $\vgamma_1$.

Note that the coefficient of $\widehat f$ in $B_1$ is an exact multiple of the matrix $T$.  So,
\begin{align*}
\footnotesize
[T,B_1] &= \dfrac{\ri}{2D^2(\Lambda_0-\lambda_0)}
\left[ \begin{pmatrix} |\rphi_2|^2 - |\rphi_1|^2 & 2\widebar\rphi_1 \widebar\rphi_2 \\
2\rphi_1 \rphi_2 & |\rphi_1|^2 - |\rphi_2|^2 \end{pmatrix},
\begin{pmatrix}
-\widebar q_1 \rphi_1\widebar\rphi_2 + q_1 \widebar\rphi_1 \rphi_2
&
q_1 \widebar\rphi_1^2 + \widebar q_1 \widebar\rphi_2^2
\\
-\widebar q_1 \rphi_1^2 - q_1 \rphi_2^2
&
\widebar q_1 \rphi_1\widebar\rphi_2 - q_1 \widebar\rphi_1 \rphi_2
\end{pmatrix} \right]
\\
&= \dfrac{\ri}{D(\Lambda_0-\lambda_0)}
\begin{pmatrix}-q_1 \widebar\rphi_1 \rphi_2 - \widebar q_1 \rphi_1 \widebar\rphi_2 & -q_1 \widebar\rphi_1^2 +\widebar q_1\widebar\rphi_2^2 \\
\widebar q_1 \rphi_1^2 -q_1 \rphi_2^2 &  -q_1 \widebar\rphi_1 \rphi_2 -\widebar q_1 \rphi_1 \widebar\rphi_2
\end{pmatrix}.
\end{align*}
Amazingly, the quantities we have to integrate are the same as before%
; in fact,
$$[T,B_1] = \dfrac{-1}{(\Lambda_0- \lambda_0)} \dfrac{\rd B_1}{\rd x}.$$
Therefore,
\begin{align*}
\vgamma_1 &=\dfrac{-1}{\Lambda_0- \lambda_0} B_1 \\
&= \dfrac{-1}{2(\Lambda_0-\lambda_0)^2D}
\left(
q_1 \begin{pmatrix} \widebar\rphi_1 \rphi_2 & \rphi_1^2 \\ -\rphi_2^2 & -\widebar\rphi_1 \rphi_2 \end{pmatrix}
+ \widebar q_1 \begin{pmatrix} -\rphi_1 \widebar\rphi_2 & \widebar\rphi_2^2 \\ -\rphi_1^2 & \rphi_1 \widebar\rphi_2\end{pmatrix}
+ \widehat f \begin{pmatrix} |\rphi_2|^2 - |\rphi_1|^2) & 2\widebar\rphi_1 \widebar\rphi_2 \\
2\rphi_1 \rphi_2 & |\rphi_1|^2 - |\rphi_2|^2\end{pmatrix}
\right)\end{align*}
When we expand this in terms of the $\Lambda_0$-natural frame of $\vgamma_0$, comprising
$T$ with
$$U_1 = \Rphi_0^{-1}\begin{pmatrix} 0 & 1 \\ -1 & 0 \end{pmatrix} \Rphi_0, \qquad
U_2 =   \Rphi_0^{-1}\begin{pmatrix} 0 & \ri \\ \ri & 0 \end{pmatrix}\Rphi_0,$$
then we obtain the  first order perturbation term in the expression for the curve
\begin{equation}\label{gamma1components}
\vgamma_1 = \dfrac{-1}{2(\Lambda_0-\lambda_0)^2}
\left(-\ri\widehat f\,T + \realpart(q_1)\,U_1 + \impart(q_1)\, U_2 \right).
\end{equation}

\subsection{Natural Frames Unlinked}
Now that we have a nice expression for the components of the perturbed
curve in terms of a natural frame, we will show that under some
circumstances the perturbed curve $\vgamma_0+\epsilon \vgamma_1$ forms a
cable around the unperturbed curve $\vgamma_0$.  To determine the type of
the cable correctly, we need to know if the natural frame itself winds
around $\vgamma_0$.  We will show that, for sufficiently small $\epsilon$,
the curve $\vgamma_0 + \epsilon U_1$ is unlinked with
$\vgamma_0$---provided that $\vgamma_0$ has self-linking number zero and
has no inflection points.  Note that, because the self-linking number is a
discrete invariant---either an integer or a half-integer---it does not
change under deformation, and is multiplied by the integer $n$ when we
pass from a curve to its $n$-fold cover.  Therefore, any curve that is
obtained from the circle by a succession of multiple-coverings and
deformations will also have zero self-linking number.

According to White's formula \cite{Wh}, the linking number of $\vgamma_0$ and $\vgamma_0+\epsilon U_1$
$$\operatorname{Lk}(\vgamma_0,\vgamma_0 + \epsilon U_1) = \operatorname{Wr}(\vgamma_0)
 + \dfrac1{2\pi}\int (T \times U_1) \rd U_1.$$
Because $\rd U_1/\rd x = -k_1 T + \Lambda_0 U_2$ (where $x$ is arclength) and $U_2 = T \times U_1$, then
$$\operatorname{Lk}(\vgamma_0,\vgamma_0 + \epsilon U_1) = \operatorname{Wr}(\vgamma_0)
 + \dfrac{L \Lambda_0}{2\pi},$$
where $L$ is the length of $\vgamma_0$.  Meanwhile, the writhe $\operatorname{Wr}(\vgamma_0)$
is related to the self-linking number by Pohl's formula \cite{Po}
$$\operatorname{SL}(\vgamma_0) =  \operatorname{Wr}(\vgamma_0)
 + \dfrac1{2\pi}\int \tau\, \rd x.$$
Taking $\operatorname{SL}(\vgamma_0)$ to be zero, we get
$$\operatorname{Lk}(\vgamma_0,\vgamma_0 + \epsilon U_1) =\dfrac1{2\pi}\int (\Lambda_0-\tau) \rd x.$$

We can relate the last integrand to the unperturbed potential
$q_0 =\tfrac12 (k_1 + \ri k_2)$ written in terms of the natural curvatures, by
making use of the Frenet equations and the natural frame equations.
Differentiating each side of the expression
$$\kappa N = k_1 U_1 + k_2 U_2$$
for the Frenet normal in terms of the natural frame vectors,
and cancelling out tangent terms on both sides gives
$$\kappa_x N + \kappa \tau B = (k_1)_x U_1 + (k_2)_x U_2 + k_1 \Lambda_0 U_2 - k_2 \Lambda_0 U_1.$$
Next, dotting both sides of the
with $\kappa B = T \times \kappa N = k_1 U_2 - k_2 U_1$ gives
$$\kappa^2 \tau = k_1 (k_2)_x- k_2(k_1)_x + \Lambda_0 \kappa^2.$$
So,
$$\Lambda_0 - \tau = -\left( \dfrac{k_1 (k_2)_x - k_2(k_1)_x}{\kappa^2} \right) = - \impart((q_0)_x/q_0).$$
Now, the value $\Lambda_0$ is chosen so that $q_0$ is periodic along the curve.  So,
provided that $q_0$ is nonvanishing along the curve (and, this is true if $\vgamma_0$
is sufficiently close to a multiply-covered circle) then the integral is zero,
and the natural frame is unlinked.

Thus, the knotting of the perturbed curve about $\vgamma_0$ is completely determined
by the behavior of $q_1$.

\subsection{Monotonicity of Argument}
In Section 4 we showed that  if $q_1$ only contains terms associated to
the double point being opened up at first order, then it must be of the form
\eqref{q1form}.  We will
establish that the argument of this function is monotone in $x$ by calculating it for
the multiply-covered circle, and invoking continuous dependence on the potential.


As an exercise, one can calculate the following Baker eigenfunction
for the plane wave solution $q=\exp(2\ri t)$ for real values of $\lambda$:
$$\vbaker(x,t;\lambda) = \dfrac{\exp(\ri k x+2\ri k\lambda t)}{k+\lambda+1}
\begin{bmatrix}\exp(\ri t)\\ (k+\lambda)\exp(-\ri t) \end{bmatrix},
$$
where $k=\sqrt{\lambda^2+1}$.  This expression extends to the spectral curve
$\mu^2 = \lambda^2+1$ associated to the plane wave solution by replacing $k$ with $\mu$,
and coincides with the expression of the Baker-Akhiezer eigenfunction normalized as in \cite{CI05}.

We then calculate that
$$c\baker_1^2 +\widebar{c}\widebar\baker_2^2 =
\exp(2\ri t)\dfrac{u + \ubar (k+\lambda)^2}{(k+\lambda+1)^2}
$$
where $u := c\exp(2\ri k x + 4\ri k\lambda t)$.
(Note that $\lambda$ and $k$ are assumed to be real.)
Because $\rd u/\rd x = 2\ri k u$, then
$$\dfrac{\rd}{\rd x} \log \left(c\baker_1^2 +\widebar{c}\widebar\baker_2^2\right)
= 2\ri k\dfrac{u - (k+\lambda)^2 \ubar}{u + (k+\lambda)^2 \ubar}.
$$
Then
\begin{multline}\label{imlogq1}
\impart\dfrac{\rd}{\rd x} \log \left(c\baker_1^2 +\widebar{c}\widebar\baker_2^2\right)
= 2k \realpart
\dfrac{u - (k+\lambda)^2 \ubar}{u + (k+\lambda)^2 \ubar}
= 2k \realpart \dfrac{(1+(k+\lambda)^4)|u|^2+ (k+\lambda)^2(u^2-\ubar^2)}{|u+(k+\lambda)^2\ubar|^2}
\\
=2k |u|^2 \dfrac{(1-(k+\lambda)^2)(1+(k+\lambda)^2)}{|u+(k+\lambda)^2\ubar|^2}
= -4k\lambda |c|^2\dfrac{(k+\lambda)(1+(k+\lambda)^2)}{|u+(k+\lambda)^2\ubar|^2},
\end{multline}
where in the last step we use the fact that $k^2 = 1+\lambda^2$ to write $1-(k+\lambda)^2 = -2\lambda(k+\lambda)$.

The sign of this quantity explains why opening up double points where $\lambda >0$ leads to left-handed
cables, and $\lambda <0$ leads to right-handed cables.  In the next section, we will determine the
type of this cable.



\subsection{Cable Type}
Given a knot $K_1:S^1 \to \R^3$ whose image lies inside a solid torus
$T^3$, and a knot $K_2:S^2 \to \R^3$, we can form a satellite knot
on $K_3 = \tau \circ K_1$, where $\tau$ is a diffeomorphism from $T^3$
to a tubular neighborhood of the image of $K_2$.  In particular,
when $K_1$ is a $(p,q)$ torus knot, then $K_3$ is a $(p,q)$ cable on $K_2$.

Because the natural frames of $\vgamma_0$ are unlinked, we can use them to define a diffeomorphism
$\delta$ from $S^1 \times D^2$ to a tubular neighborhood of $\vgamma_0$,
given by
$$\delta:(z,r,\theta) \mapsto \vgamma_0(z) + (r\cos \theta) U_1(z) + (r\sin \theta) U_2(z),$$
where $z,r,\theta$ are cylindrical coordinates, with $r$ less than one half of the minimum radius
of curvature of $\vgamma_0$, and
we think of $S^1 \times D^2$ as the quotient of the solid cylinder
by the equivalence relation $z \sim z+L$.


By \eqref{gamma1components}, the perturbed curve is
$$\vgamma(x) = \vgamma_0(x) -
\dfrac{\epsilon}{2(\Lambda_0-\lambda_0)^2}
\left(-\ri \widehat f\, T + \realpart(q_1)\,U_1 + \impart(q_1)\, U_2 \right)
+O(\epsilon^2).$$
Consider the modification
$$\widetilde\vgamma(x) =\vgamma_0(x) -\dfrac{\epsilon}{2(\Lambda_0-\lambda_0)^2}
\left(\realpart(q_1)\,U_1 + \impart(q_1)\, U_2 \right)
,$$
which, by continuity, will have the same knot type as $\vgamma(x)$ for $\epsilon$ sufficiently small.
Because $q_1$ is composed of quadratic products of eigenfunctions, which have the
form \eqref{bakerform}, the period of $q_1$ is the lowest common multiple of $L=n\pi$ and
$\pi/\Omega_1(\lambda_0)$.  Because the isoperiodic deformations preserve the value of
$\Omega_1(\lambda_0)$, we can compute it by running the deformations backwards to the
multiply-covered circle.
Specifically, suppose that $[n;m_1, \ldots, m_g]$ is the deformation scheme that produced $q_0$.
Then, upon running the deformation steps backwards,
the basepoint for $\Omega_1$ limits to $-\ri$.  So,
$$\Omega_1(\lambda_0) = \int_{-\ri}^{\sqrt{(m/(jn))^2-1}} \dfrac{\lambda}{\sqrt{\lambda^2+1}}\,\rd \lambda = \dfrac{|m|}{jn}.$$
Hence, $\pi/\Omega_1(\lambda_0) = j L /|m|$, and the period of $q_1$ is $j L$.  (Recall that $j$ and $m$ are coprime.)

The image of $\widetilde\vgamma(x)$ under the inverse of $\delta$ is
the curve
$$z=x, \qquad r=\epsilon |q_1(x)|/(2(\Lambda_0-\lambda_0)^2), \qquad \theta = \arg(-q_1(x)).$$
Note that $r$ is never zero along this curve, so that it never crosses
the $z$-axis.  Consider the map $\overline{\theta}:S^1 \to S^1$ defined by
$x \mapsto \arg(q_1(x))$, where the domain $S^1$ is $\R$ modulo $jL$ and the
codomain $S^1$ is $\R$ modulo $2\pi$.  For a fixed $c$,
$$q_1=c\baker_1^2 + \widebar{c}\,\widebar\baker_2^2 = c\left(\baker^+_1(q_0,\lambda_0)\right)^2 + \widebar{c}(\widebar\baker^+_2(q_0,\lambda_0))^2$$
 depends continuously on the
location of $\lambda_0$ and the potential $q_0$.  Thus, the degree of the mapping
$\overline{\theta}$ is unchanged under the multi-step isoperiodic deformation process.  We can calculate
its degree by substituting $k=-m/(jn)$ and $\lambda=-\operatorname{sign}(m)\sqrt{k^2+1}$ into formula
\eqref{imlogq1}.  Integrating that formula shows that
\begin{equation}\label{badintegral}
\impart\,\log\left(c\baker_1^2 + \widebar{c}\,\widebar\baker_2^2\right)
= - \frac{\ri}{2}\log\left(\dfrac{u + (k+\lambda)^2\ubar}{\ubar + (k+\lambda)^2 u}\right).
\end{equation}
So, the period of $\arg(q_1)$ is $\pi/|k| = j n \pi/|m|$.  Therefore, the degree of mapping $\overline{\theta}$ is $m$,
taking the minus sign in \eqref{badintegral} into account.

Because the curve $\delta^{-1}(\widetilde\vgamma(x))$ winds around the $z$-axis $|m|$ times (in a counterclockwise
direction when $m>0$ and clockwise when $m<0$) we see that it is a $(j,m)$ torus knot in $S^1 \times D^2$.
Thus, the curve $\widetilde\vgamma(x)$ is a $(j,m)$ cable about $\vgamma_0$.

\subsection{Conclusions.}  We may now state our main result in full:
\begin{theorem}
\label{cabletheorem_formal}
Let $n,m_1,\ldots m_K$ be relatively prime with $|m_k|>n>0$, and let $g_k = \gcd(n,m_1,\ldots,m_k)$.
Let $(*)_k$ be the isoperiodic deformation system with variables
$\alpha_\ell$, $\lambda_j$ for $0 \le \ell \le K$, $1\le j \le 2K+2$,
controls
$c_k = 1$, $c_\ell=0$ for $\ell \ne k$, and with change of variable
$\xi = \tfrac12 \epsilon^2$.  Then:

1. For each $k$ between 1 and $K$ there exists $\rho_k >0$ such that
$(*)_k$ has analytic solution $(\alpha^{(k)}_\ell,\lambda^{(k)}_j)$
for $|\epsilon|<\rho_k$ satisfying, when $k=1$,
\begin{gather*}
\alpha^{(1)}_0(0) = 0,\quad \lambda^{(1)}_1(0) = \ri, \quad\lambda^{(1)}_2(0) = -\ri,
\dfrac{d \lambda^{(k)}_{3}}{d\epsilon} (0) =
-\dfrac{d \lambda^{(k)}_{4}}{d\epsilon} (0) = \ri,\\
\alpha^{(1)}_\ell(0) = \lambda^{(1)}_{2\ell+1}(0) = \lambda^{(1)}_{2\ell+2}(0)
= -\sign(m_\ell)\sqrt{(m_\ell/n)^2-1}, \quad 1\le \ell \le K.
\end{gather*}
and when $k>1$ (with $\rho_k$ depending on the choice of $\epsilon_{k-1} \in (0,\rho_{k-1})$)
$$
\alpha^{(k)}_\ell(0) = \alpha^{(k-1)}_\ell(\epsilon_{k-1}),
\lambda^{(k)}_j(0) = \lambda^{(k-1)}_j(\epsilon_{k-1}),
\dfrac{d \lambda^{(k)}_{2k+1}}{d\epsilon} (0) =
-\dfrac{d \lambda^{(k)}_{2k+2}}{d\epsilon} (0) = \ri
$$

2. For each $k$ there exist finite-gap potentials $q^{(k)}(x;\epsilon)$  which are $n\pi$-periodic in $x$,
analytic in $\epsilon$, and
for which the simple points are
$\lambda^{(k)}_j(\epsilon)$, $1\le j \le 2k+2$.
Then the filament $\vgamma^{(k)}(x;\epsilon)$, constructed
from $q^{(k)}$ using the Sym-Pohlmeyer formula at $\Lambda_0=\alpha^{(k)}_0(\epsilon)$,
is closed of length $\pi n/g_k$ and is a $(g_{k-1}/g_k, m_k/g_k)$-cable
about $\vgamma^{(k)}(x;\epsilon_k)$.
\end{theorem}

The argument given in Section \ref{isogap} implies that for each $k$ there is a
deformation $q^{(k)}(x,t;\epsilon)$ of NLS {\em solutions} whose spectrum matches the
deformation of the branch points at any time.  The above theorem,
applied at any fixed time $t_0$, implies that the corresponding filament
$\vgamma^{(k)}(x,t_0;\epsilon_k)$ has the desired cable type.  (Note that the
plane wave potential $q^{(0)}$ at time $t_0$ only differs from that
at time zero by multiplication by a unit modulus constant.)
Thus, we have the

\begin{corollary}\label{timecorollary}
The knot type of $\vgamma^{(k)}(x,t;\epsilon_k)$ is fixed for all time.
\end{corollary}

The local nature of the VFE can, in general, cause the knot type to
change in time.  Indeed, it is relatively easy to construct
solutions with changing knot type by taking B\"acklund transformations
of genus one finite-gap solutions.  Thus the Corollary
implies that we can nonetheless construct a neighborhood of the
multiply-covered circle, within the class of finite-gap VFE solutions,
which consists of filaments whose knot is preserved.

%% file: isoapp1.tex
 \section{Appendix: Completeness of the Squared Eigenfunctions}
 \label{isoapp1}

In this Appendix we summarize the main properties of squared
eigenfunctions for the NLS equation, in particular their connection to the
linearized NLS equation, their biorthogonality property, and the
$L^2$-completeness of a suitably chosen periodic subfamily.

We rewrite the NLS equation as a dynamical system
\begin{equation}
\label{NS2}
\begin{split}
\ri q_t & = - q_{xx} + 2q^2r, \\
\ri r_t & = r_{xx} - 2r^2 q,
\end{split}
\end{equation}
for the pair  $(q,r)$ in the phase space $\mathcal{P}=H^1_{per}([0,L],\mathbb{C}^2)\subset L^2([0,1], \mathbb{C}^2)$ of periodic, square integrable, vector-valued functions of $x$, with square integrable first  derivative. The inner product is taken to be the one of the ambient space:
\begin{equation}
\label{innerp}
\left\langle \mathbf{f},\mathbf{g} \right\rangle=\frac{1}{L}\int_0^L [f_1(x)\bar{g}_1(x)+f_2(x)\bar{g}_2(x)]\, \rd x,
\end{equation}
where $\mathbf{f}=(f_1, f_2)$, $\mathbf{g}=(g_1, g_2)$ are in $\mathcal{P}$.
\smallskip

The focusing/defocusing reality condition in achieved by restricting  to one of the invariant subspaces $\mathcal{P}^\pm=\{ (q,r)\in \mathcal{P} \, | \, r=\pm \bar{q} \}$ on which the inner product \eqref{innerp} becomes the real inner product
\[
 \left\langle\mathbf{f}, \mathbf{g}\right\rangle_R=\frac{2}{L} \realpart{\left( \int_0^L u(x)\bar{v}(x) \, \rd x \right),}
 \]
 where $\mathbf{f}=(u, \pm \bar{u})$, $\mathbf{g}=(v, \pm \bar{v})$ are elements of $\mathcal{P}^\pm$.

For a given NLS potential $\mathbf{q}=(q,r)\in \mathcal{P}$, we consider the linearization of the NLS system  \eqref{NS2}, obtained by replacing $q\rightarrow q+u$, $r\rightarrow r+v$ and retaining terms up to first order in $u$ and $v$,
\begin{equation}
\label{LNS2}
\begin{split}
\ri u_t + u_{xx} - 2q^2v-4qru & = 0 \\
\ri v_t -v_{xx} +2r^2 u + 4 qr v & =0.
\end{split}
\end{equation}
For fixed time $t_0$, we regard the pair $(u,v)$ as an element of the ambient space $L^2([0,L], \mathbb{C}^2)$, endowed with Hermitian inner product given by \eqref{innerp}.

\subsection{Squared Eigenfunctions}\label{squaredapp}
If ${\vphi}$ and ${\widetilde{\vphi}}$ are solutions of the NLS linear system \eqref{LINS} at the same value of $\lambda$, then the triplet
\begin{equation}
\label{SQEIG}
{\vphi}\otimes {\widetilde{\vphi}}=\left[ \begin{array}{c} \frac{1}{2} (\phi_1\widetilde{\phi}_2 + \widetilde{\phi}_1\phi_2) \\ \phi_1\widetilde{\phi}_1 \\ - \phi_2 \widetilde{\phi}_2 \end{array} \right]=\left[
\begin{array}{c}  f \\ g \\ h \end{array} \right],
\end{equation}
solves the {\sl squared eigenfunction} systems:

\begin{equation}\label{fghxsys}
\begin{bmatrix}f \\ g\\ h \end{bmatrix}_x =
\begin{pmatrix} 0 & -\ri r& -\ri q \\
2 \ri q & -2\ri \lambda & 0 \\
2\ri r & 0 & 2\ri \lambda \end{pmatrix}
\begin{bmatrix} f \\ g\\ h \end{bmatrix}.
\end{equation}

\begin{equation}\label{fghtsys}
\begin{bmatrix}f \\ g\\ h \end{bmatrix}_t =
\begin{pmatrix} 0 & -(2\ri\lambda r + r_x)& - (2\ri\lambda q -q_x) \\
2(2\ri\lambda q -q_x)& -2(2\ri\lambda^2 + \ri q r) & 0 \\
2(2\ri\lambda r +  r_x)& 0 & 2(2\ri\lambda^2+\ri q r) \end{pmatrix}
\begin{bmatrix} f \\ g\\ h \end{bmatrix}.
\end{equation}

The following properties can be easily verified using systems \eqref{fghtsys}, \eqref{fghxsys}:
\begin{proposition}
(1) (linearization)
The pair $(g,h)=(\phi_1\widetilde{\phi}_1, -\phi_2\widetilde{\phi}_2)$ solves the linearized NLS equations \eqref{LNS2}.
\\
(2) (biorthogonality)
Suppose that $\lambda \not= \mu$ are distinct points of the periodic/antiperiodic spectrum of the spatial linear operator $\displaystyle \mathcal{L}_1=\ri\sigma_3\frac{d}{d x} +\left( \begin{array}{cc} 0 & q \\ r & 0 \end{array} \right)$. Let $(g(\lambda),h(\lambda))$ and $(\widetilde{g}(\mu), \widetilde{h}(\mu))$  be solutions of the squared eigenfunction systems. Then,
\begin{equation}
\label{orth}
\left\langle\left( \begin{array}{c}g(\lambda) \\ h(\lambda) \end{array}\right), J \overline{ \left( \begin{array}{c}\widetilde{g}(\mu) \\ \widetilde{h}(\mu) \end{array}\right)}\right\rangle=0.
\end{equation}
\end{proposition}

\subsection{A basis of squared eigenfunctions}

Given a finite-genus NLS potential $q$, one can construct  a periodic subfamily of  solutions of the linearized NLS equation which, for fixed $t$, is also a basis for $L^2([0,L], \mathbb{C}^2)$. Its elements are squared eigenfunctions constructed from Bloch eigenfunctions evaluated at a countable number of points of the spectrum of $q$.

The Bloch eigenfunctions are common solutions of the AKNS system and the shift operator, i.e., they satisfy the additional property
\[
\boldsymbol{\varphi}^\pm(x+L)=\rho^\pm(\lambda)\boldsymbol{\varphi}^\pm(x),
\]
where $\rho^\pm(\lambda)$ are the Floquet multipliers
\begin{equation}
\label{FloquetM}
\rho^\pm(\lambda) = \frac{\Delta(\lambda)\pm \sqrt{\Delta^2(\lambda)-4}}{2}.
\end{equation}
A useful normalization is obtained by selecting (see \cite{LM, MOv})
\begin{equation}
\label{normalizedbloch}
\begin{aligned}
\boldsymbol{\varphi}^+ & =e^{\ri\frac{\pi}{4}} \sqrt{\frac{\rho^-(\lambda)-M_{11}(\lambda)}{M_{12}(\lambda)(\Delta^2(\lambda)-4)}} \left[ M_{12}(\lambda) \mathbf{Y}_1+(\rho^+(\lambda)-M_{11}(\lambda)) \mathbf{Y}_2 \right], \\
\boldsymbol{\varphi}^- & =e^{\ri\frac{\pi}{4}} \sqrt{\frac{\rho^+(\lambda)-M_{11}(\lambda)}{M_{12}(\lambda)(\Delta^2(\lambda)-4)}} \left[ M_{12}(\lambda) \mathbf{Y}_1+(\rho^-(\lambda)-M_{11}(\lambda)) \mathbf{Y}_2 \right].
\end{aligned}
\end{equation}
so as to satisfy the useful symmetry
\[
\boldsymbol{\varphi}^-(x;\lambda)=J\overline{\boldsymbol{\varphi}^+(x;\bar{\lambda})}.
\]
In the expressions above, $M_{ij}(\lambda)$ are the entries of the transfer matrix $M(x,t;\lambda)=\Phi(x+L, t;\lambda)\Phi(x, t;\lambda)^{-1}$ across a spatial period $L$, and the vectors $\mathbf{Y}_i$ are the columns of the fundamental matrix solution of the AKNS system \eqref{LNS2}.

For simplicity, we consider an NLS potential $(q,r)$ the critical points of which are of algebraic and geometric multiplicity two. (Higher order critical points introduce some technical difficulty, while critical points of geometric multiplicity one can be dealt with in a simple way.)
We define the following family of squared eigenfunctions.

\noindent
(A) At the real and complex double points $\{ \lambda_j^d\}$:
\[
\mathbf{f}^{(+,j)}(x)=\left( \begin{array}{c} (\varphi^+_1)^2(x;\lambda) \\ -(\varphi^+_2)^2(x;\lambda)\end{array}\right)_{\lambda=\lambda_j^d}, \quad \mathbf{f}^{(-,j)}=\left( \begin{array}{c} (\varphi^-_1)^2(x;\lambda) \\ -(\varphi^-_2)^2(x;\lambda) \end{array}\right)_{\lambda=\lambda_j^d}.
\]
(B) At the finite number $N$ of critical points $\{\lambda^c_k\}$ that are not elements of the periodic/antiperiodic spectrum of $q$:
\[
\mathbf{f}^{(c,k)}=\left( \begin{array}{c} (\varphi^+_1 \varphi^-_1)(x;\lambda) \\ -(\varphi^+_2\varphi^-_2)(x;\lambda)\end{array}\right)_{\lambda=\lambda_k^C},
\]
(C) At $N$ (appropriately selected) simple points $\{\lambda^s_i\}$ of the spectrum (half the number of branch points of the associated Riemann surface):
\[
\mathbf{f}^{(s,i)}=\left( \begin{array}{c} (\varphi^+_1)^2(x;\lambda) \\ -(\varphi^+_2)^2(x;\lambda)\end{array}\right)_{\lambda=\lambda_{2i}^S}, \quad k, i=1, \ldots N.
\]
Then
\begin{theorem}[Completeness]
The family
\[
\left\{ \mathbf{f}^{(+,j)}, \mathbf{f}^{(-,j)} \right\}_{\lambda_j^d} \bigcup \left\{ \mathbf{f}^{(c,k)} \right\}_{\lambda^c_k}
  \bigcup \left\{ \mathbf{f}^{(s,i)}  \right\}_{\lambda^s_i}
 \]
 is a basis of $L^2([0,L], \mathbb{C}^2)$.
\end{theorem}
A proof of this fact can be found in the preprint \cite{CE06}, following the roadmap   developed in \cite{EFM1} for the sine-Gordon equation and incorporating ideas from \cite{Kr92}.
We also mention the following result:
\begin{lemma}[Biorthogonal Pairing]
\label{biopair}
The  families  $\{\mathbf{f}^{(+,j)}\}$ and  $\{J\overline{\mathbf{f}^{(-,k)}}\}$;
 $\{\mathbf{f}^{(-,j)}\}_{|j|>N}$ and  $\{J\overline{\mathbf{f}^{(+,k)}}\}$,
are biorthogonally paired with respect to the Hilbert space inner product \eqref{innerp}:
\[
\left\langle\mathbf{f}^{(\pm,j)},J\overline{\mathbf{f}^{(\mp,k)}}\right\rangle=C_j\delta_{jk},
\]
where $\displaystyle C_j=\frac{1}{2\ri} \sqrt{\Delta(\lambda_j)\Delta''(\lambda_j)} \left( W[\boldsymbol{\varphi}^+(x; \lambda_j), \boldsymbol{\varphi}^-(x; \lambda_j)]\right)^2.$
\end{lemma}
Notice that the existence of a biorthogonal pairing automatically guarantees that the family of squared eigenfunctions evaluated at the double points of the spectrum of $q$ form a linearly independent set.

%% file: isogap.tex
\section{Appendix: Analytic Dependence of the Potential on the Deformation Parameter}\label{isogap}
The reconstruction of a finite-gap potential of genus $g$ relies on choice of
$2g+2$ branch points in the complex plane and also on a choice of an effective divisor $\cD$ of degree $g+1$ (satisfying a certain
reality condition) on the hyperelliptic Riemann surface defined by the branch points.  Thus, to specify an isoperiodic
deformation of the NLS potential $q$ we must specify not only a deformation
of the branch points but also a deformation of the divisor (or an equivalent
set of auxiliary data).  In this appendix, we show that, at the same time that
a double point is being opened up by homotopic deformation,
the auxiliary data given by the Dirichlet spectrum may be deformed in a way that guarantees that
$q$ is analytic in the deformation parameter $\epsilon$.  This result is necessary for us to carry
out the perturbation expansions in \S\ref{isoperturb} and \S\ref{isocabling}.

\subsection{Finite-Gap Solutions and the Dirichlet Spectrum}


The Dirichlet spectrum of an NLS solution $q(x,t)$ which is $L$-periodic in $x$
is defined as follows (see \cite{AMa}, \cite{MOv}).  Given a fundamental
matrix solution $\Rphi$ for the AKNS system
\begin{equation}
\label{LINS}
\vrphi_x = \begin{pmatrix} -\ri \lambda & \ri q \\ \ri \bar{q} & \ri \lambda
\end{pmatrix}\vrphi,
\quad
\vrphi_t = \begin{pmatrix}\ri (|q|^2-2\lambda^2) & 2\ri \lambda q - q_x \\
2\ri \lambda \bar{q} +\bar{q}_x & -\ri (|q|^2-2\lambda^2)\end{pmatrix}\vrphi,
\end{equation}
we construct the transfer matrix
$$M(x,t;\lambda) = \Rphi(x+L,t;\lambda) \Rphi(x,t;\lambda)^{-1},$$
There are two sets of Dirichlet eigenvalues:
\begin{itemize}
\item The Dirichlet eigenvalues $\mu_k(x,t)$ are the $\lambda$-values for which
\begin{equation}\label{mucond}
M_{11}-M_{22}+M_{12}-M_{21}=0.
\end{equation}
Equivalently, these are the values for which there is a nontrivial
solution of \eqref{LINS} satisfying the boundary condition $\rphi_1 +\rphi_2 = 0$
at $(x,t)$ and at $(x+L,t)$.
\item
The Dirichlet eigenvalues $\nu_k(x,t)$ are the $\lambda$-values for which
$$M_{11}-M_{22}-\ri(M_{12}+M_{21})=0.$$
Equivalently, these are the values for which
there is a nontrivial solution of \eqref{LINS} satisfying $\ri \rphi_1 + \rphi_2=0$
at $(x,t)$ and $(x+L,t)$.  Also, these are the $\mu$-eigenvalues that result when $q$ is replaced by $\ri q$.
\end{itemize}
The $\mu_k$ and $\nu_k$ are dependent on $x$ and $t$, satisfying a system
of ODEs first derived by Ablowitz and Ma \cite{AMa}.  However, for finite-gap
solutions, all but finitely many of each set are locked to the double points of the
Floquet spectrum.  (This is true more generally for periodic NLS potentials whose
Sobolev norm over one period is bounded; see \cite{MOv}.)

In order to get the equations defining the Dirichlet eigenvalues, we need
to use the formulas for finite-gap NLS solutions and their associated
Baker eigenfunction.  We will briefly review their construction; further
details are available in the book \cite{BBEIM} or our paper \cite{CI05}.

\begin{figure}[h]
\centering
\includegraphics[height=2.4in]{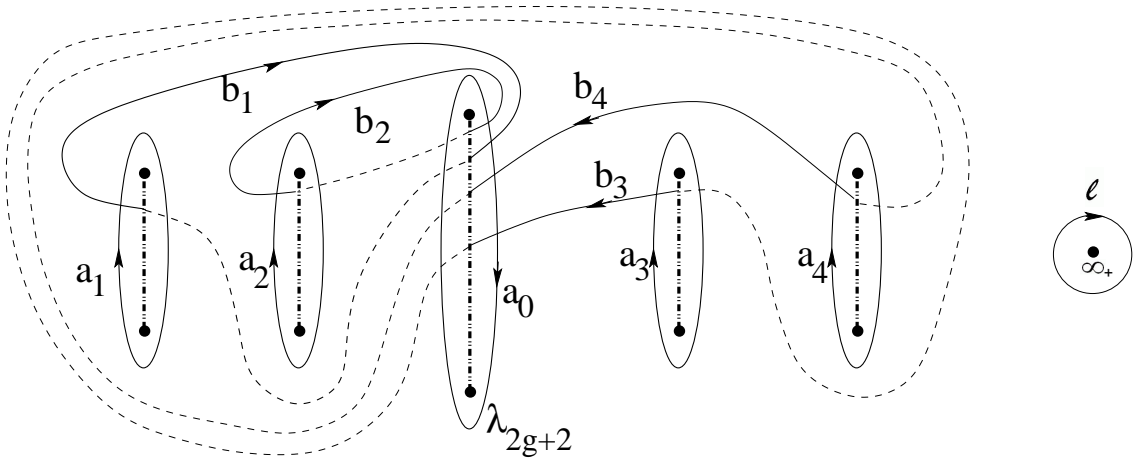}
\caption{Homology basis adapted to isoperiodic deformations, for genus $g=4$.
Branch cuts extend between complex conjugate branch points; solid paths lie
on the upper sheet, dashed on the lower sheet; basepoint for Abelian
differentials $\Omega_1,\Omega_2,\Omega_3$ is marked as $\lambda_{2g+2}$.
The extra cycles $a_0$ and $\ell$ are used in the proof below.}\label{althbasis}
\end{figure}

One starts with a hyperelliptic
Riemann surface $\Sigma$ defined by
\begin{equation}\label{Riemannsurf}
\zeta^2 = \prod_{j=1}^{2g+2} (\lambda-\lambda_j),
\end{equation}
where the branch points $\lambda_j$ occur in complex conjugate
pairs.  This surface is compactified by the points $\infty_\pm$, where
$\lambda^{g+1}/\zeta \to \pm 1$.
After selecting a suitable homology basis $a_j,b_j$ (see Figure \ref{althbasis}),
one constructs holomorphic differentials $\omega_j$ such that $\oint_{a_k} \omega_j = 2\pi \ri \delta_{jk}$
and meromorphic differentials $\rd\Omega_1,\rd\Omega_2,\rd\Omega_3$ with prescribed singularities
at $\infty_\pm$ and zero $a$-periods.  Then
\begin{equation}\label{finiteq}
q(x,t) = A \exp(-\ri E x + \ri N t) \dfrac{\theta(\ri\vV x +\ri \vW t - \vD - \vr)}{\theta(\ri\vV x +\ri \vW t - \vD)},
\end{equation}
where
\begin{itemize}
\item $\theta:\bC^g \to \bC$ is the Riemann theta function, with period lattice generated
by the vectors $2\pi \ri \ve_j$ (where $\{\ve_j\}$ is the standard basis of unit vectors)
and with quasiperiods given by the columns of the Riemann matrix $B$ defined
by $B_{jk} = \oint_{b_k} \omega_j$;
\item $\vV,\vW$,and $-\vr$ are the vectors of $b$-periods of $\rd\Omega_1,\rd\Omega_2,\rd\Omega_3$ respectively;
\item the Abelian integrals $\Omega_1,\Omega_2,\Omega_3$, with the branch point $\lambda_{2g+2}$
in the lower half-plane as basepoint, have asymptotic behavior
\begin{align*}
\Omega_1(P) &\sim \pm \left(\lambda - \frac{E}{2} + o(1)\right),
\\
\Omega_2(P) &\sim \pm \left(2\lambda^2 + \frac{N}{2} + o(1)\right),
\\
\exp(\Omega_3(P)) &\sim \pm\left(\frac{2\ri}{A} \lambda + o(1) \right)
\end{align*}
as $P \to \infty_\pm$, where $A$ is real and positive.  (Because they have nonzero $b$-periods, the integrals
are not well-defined on $\Sigma$, but are well-defined on the surface $\Sigma_0$
obtained after cutting $\Sigma$ along the homology basis.)
\item The vector $\vD$ is defined so that the zero divisor
of $\theta(\A(P) - \vD)$ is the positive divisor $\cD_+$ linearly equivalent
to $\cD - \infty_+$, where $\A$ is the Abel map defined by
$$\A(P) = \int_{\infty_-}^P \vomega$$
and $\vomega$ is the vector of holomorphic differentials $\omega^j$.
(The reality condition on $\cD$ is that $\vD$ have zero real part.)
\end{itemize}
A vector solution of \eqref{LINS} is provided at each $P\in \Sigma$ by
the Baker eigenfunction $\vbaker$, with components
$$\baker_1(x,t) = \dfrac{\exp\left(\ri (\Omega_1(P)x+\Omega_2(P)t)\right)}{\exp(\ri(E x-N t)/2)}
\dfrac{\theta(\A(P) + \ri\vV x +\ri \vW t - \vD)\theta(\vD)}
{\theta(\ri\vV x +\ri \vW t - \vD)\theta(\A(P)-\vD)}
$$
and
$$\baker_2(x,t) =-\ri \dfrac{\exp\left(\ri (\Omega_1(P)x+\Omega_2(P)t)+\Omega_3(P)\right)}{\exp(\ri(N t-E x)/2)}
\dfrac{\theta(\A(P) + \ri\vV x +\ri \vW t - \vD-\vr)\theta(\vD)}
{\theta(\ri\vV x +\ri \vW t - \vD)\theta(\A(P)-\vD)}.
$$
Although $\A(P)$ and $\Omega_i(P)$ are not well-defined on $\Sigma$ because of nontrivial
periods, we make the convention that the paths of integration for $\A(P)$ and $\Omega_i(P)$ differ
by a fixed path from $\infty_-$ to $\lambda_{2g+2}$ in the cut surface $\Sigma_0$, and this
makes the Baker eigenfunction well-defined.  (The rightmost theta-factors in the numerator
and denominator, which were unfortunately omitted in \cite{CI05}, are necessary for this.)

Let $\interchange:(\lambda,\zeta)\mapsto (\lambda,-\zeta)$ be the sheet interchange involution of the
Riemann surface.  When $P$ is not a branch point, the Baker eigenfunctions $\vbaker(P)$ and $\vbaker(\interchange P)$ are a basis for
solutions of \eqref{LINS}.  Thus, to compute the initial values $\mu_k(0,0)$ and $\nu_k(0,0)$ of
the Dirichlet spectrum, we evaluate $\vbaker$ when $t=0$ and $x=0$ or $x=L$.  Using the facts
that $L\vV/(2\pi)$ and $LE/(2\pi)$ are integer-valued, we get
\begin{align*}
\baker_1(0,0)& =1,& \baker_2(0,0)&=-\ri f(P), \\
\baker_1(L,0) &= \exp\left(\ri L\left(\Omega_1(P)-\frac{E}2\right)\right), &
\quad \baker_2(L,0) &= -\ri f(P) \baker_1(L,0),
\end{align*}
where
$$f(P) = \exp(\Omega_3(P))\dfrac{\theta(\A(P)-\vD-\vr)}{\theta(\A(P)-\vD)}.$$
So, $\baker_1 + \baker_2=0$ is satisfied at both ends if and only either $\sin(L\Omega_1(P))=0$ or
$f(P)=-\ri$.  This reflects the fact that every double point of the Floquet spectrum
is a Dirichlet eigenvalue which is constant in $x$ and $t$.  On the other hand,
the function $f(P)$ is well-defined and meromorphic on $\Sigma$, with pole divisor $\cD_+ +\infty_+$,
so there are $g+1$ Dirichlet eigenvalues $\mu_k(0,0)$ that are not locked to the double points, and are
located at the images of the points where $f(P)=-\ri$ under the projection $\pi:\Sigma \to \bC$
onto the complex $\lambda$-plane.
Similarly, the $\nu_k(0,0)$ are either locked to double points or are images of the $g+1$ points where $f(P)=1$
when $x=t=0$.

\subsection{Deformations maintaining $\vD=0$}

\begin{prop}\label{Dnuprop}  When $\vD=0$, the points where $f(P) = 1$ (and hence the locations of $\nu_k(0,0)$)
are $\lambda_{2g+2}$ and the
branch points in the upper half plane (excluding $\overline{\lambda_{2g+2}}$).
At the other branch points, $f(P)=-1$.
\end{prop}

Before sketching the proof, we note that in order for the $\nu$-eigenvalues
to deform continuously when further double points are opened up, the basepoint
$\lambda_{2g+2}$ must change continuously.  Thus, we cannot take the
convention (which we used in \cite{CI05}) that it be located at the rightmost
branch point in the lower half plane.  Instead, we will keep it located
at the branch point in the lower half plane that was originally part of
the plane wave spectrum, located on the negative imaginary axis.
Because this basepoint must be in the interior of the cut Riemann surface,
we must rearrange our homology basis so that its cycles do not
enclose this basepoint.  (The extra cycle $a_0$ in the following diagram
is not part of the homology basis.)

\begin{proof}
Because $\lambda_{2g+2}$ is the basepoint for $\Omega_3$ and $\A(\lambda_{2g+2})=\vr/2$,
$f(\lambda_{2g+2})=1$.  Let $\lambda_{2g+1}=\overline{\lambda_{2g+2}}$ in the
upper half plane, and fix an integration path from $\lambda_{2g+2}$ to
$\lambda_{2g+1}$ on the upper sheet on the right side of the branch cut.
Because $\rd\Omega_3$ is a linear combination of differentials of the
form $(\lambda^k/\zeta) d\lambda$, then $\Omega_3(\lambda_{2g+1}) = -\frac12 \oint_{a_0} \rd\Omega_3$.
Furthermore,  because $a_0+a_1 + \ldots + a_g$ is
homologous to $-\ell$ in the surface $\Sigma$ with the points $\infty_\pm$
removed, then using the residue of $\rd\Omega_3$ at $\infty_+$ we get
$\Omega_3(\lambda_{2g+1}) = \frac12 \oint_{\ell}\rd\Omega_3 = \pi \ri.$
Similarly,
$$\A(\lambda_{2g+1}) = \A(\lambda_{2g+2}) - \frac12 \oint_{a_0} \vomega =
\frac12 \vr + \pi \ri \vone.$$
where $\vomega$ is the vector of normalized holomorphic differentials $\omega_k$,
and $\vone$ is the vector whose entries are all equal to 1.  Then, using
the even-ness and periodicity of $\theta$,
$$f(\lambda_{2g+1}) = \exp(\pi\ri)
\dfrac{\theta(\A(\lambda_{2g+1})-\vr)}{\theta(\A(\lambda_{2g+1}))}
=-\dfrac{\theta(\pi\ri\vone - \tfrac12 \vr)}{\theta(\pi\ri\vone+\tfrac12 \vr)}=
-\dfrac{\theta(\tfrac12 \vr - \pi \ri \vone)}{\theta(\pi\ri\vone+\tfrac12 \vr)}
=-1.$$

Suppose $\realpart(\lambda_{2k-1}) <\realpart(\lambda_{2g+2})$
and $\impart(\lambda_{2k-1})>0$.  Let $\widetilde{b_k}$ be the cycle
that runs from $\lambda_{2k-1}$ to $\lambda_{2g+1}$ along the upper sheet,
and back along the same path along the lower sheet, whose projection
to the $\lambda$-plane encircles no other branch points.  Then
$\widetilde{b_k} \sim b_k + \sum a_m$,
where the sum is over the $a$-cycles around branch cuts whose real part
is strictly between that of $\lambda_{2k-1}$ and $\lambda_{2g+1}$.
Letting the path of integration from $\lambda_{2g+2}$ to $\lambda_{2g+1}$
be continued by the part of $\widetilde{b_k}$ on the lower sheet,
$$\Omega_3(\lambda_{2k-1}) = \Omega_3(\lambda_{2g+1})+ \tfrac12\oint_{\widetilde{b_k}}\rd \Omega_3
= \pi\ri -\tfrac12 r_k$$
and
$$\A(\lambda_{2k-1}) = \A(\lambda_{2g+1}) + \tfrac12 \oint_{\widetilde{b_k}}\rd\Omega_3
=\tfrac12(\vr+B_k) + \pi\ri (\vone + \sum \ve_m),$$
where $B_k$ indicates the $k$th column of the Riemann matrix.  Then using
evenness and periodicity of $\theta$,
$$\dfrac{\theta(\A(\lambda_{2k-1})-\vr)}{\theta(\A(\lambda_{2k-1}))}=
\dfrac{\theta(\tfrac12(-\vr+B_k) + \pi\ri (\vone + \sum \ve_m))}
{\theta(\tfrac12(\vr + B_k) + \pi\ri (\vone + \sum \ve_m))}
= \dfrac{\theta(\tfrac12(\vr-B_k) + \pi\ri (\vone + \sum \ve_m))}
{\theta(\tfrac12(\vr - B_k)  + \pi\ri (\vone + \sum \ve_m)+B_k)}.
$$
Using the quasiperiodicity property $\theta(\vz + B_k) = \exp(-z_k - \tfrac12 B_{kk})\theta(\vz)$,
$$f(\lambda_{2k-1}) = \dfrac{\exp(\pi\ri -\tfrac12 r_k)
\theta(\tfrac12(\vr-B_k) + \pi\ri (\vone + \sum \ve_m))}
{\exp(-\tfrac12(r_k-B_{kk})+\pi \ri -\tfrac12 B_{kk})\theta(\tfrac12(\vr - B_k) + \pi\ri (\vone + \sum \ve_m))}
=1.$$
Let $\lambda_{2k}=\overline{\lambda_{2k-1}}$ in the lower half-plane, and let
the path from $\lambda_{2g+2}$ to $\lambda_{2k-1}$ be continued
by the part of $a_k$ to the left of the branch cut.  Then
$\Omega_3(\lambda_{2k}) = \Omega_3(\lambda_{2k-1}) - \tfrac12 \oint_{a_k}\rd\Omega_3 =
\Omega_3(\lambda_{2k-1})$ and
$\A(\lambda_{2k}) = \A(\lambda_{2k-1}) - \pi \ri \ve_k$.
Using evenness and periodicity,
$$\dfrac{\theta(\A(\lambda_{2k})-\vr)}{\theta(\A(\lambda_{2k}))}
=\dfrac{\theta(\tfrac12(-\vr+B_k)+\pi\ri(\vone - \ve_k + \sum \ve_m))}
{\theta(\tfrac12(\vr+B_k)+\pi\ri(\vone - \ve_k + \sum \ve_m))}
=\dfrac{\theta(\tfrac12(\vr-B_k)+\pi\ri(\vone - \ve_k + \sum \ve_m))}
{\theta(\tfrac12(\vr-B_k)+\pi\ri(\vone - \ve_k + \sum \ve_m)+B_k)}.$$
Then, using quasiperiodicity,
$$f(\lambda_{2k}) = \dfrac{\exp(\pi\ri -\tfrac12 r_k)
\theta(\tfrac12(\vr-B_k) + \pi\ri (\vone -\ve_k+ \sum \ve_m))}
{\exp(-\tfrac12(r_k-B_{kk}) -\tfrac12 B_{kk})\theta(\tfrac12(\vr - B_k) + \pi\ri (\vone -\ve_k+ \sum \ve_m))}
=-1.$$

Similarly, assuming that
$\realpart(\lambda_{2k-1})>\realpart(\lambda_{2g+1})$, $\impart(\lambda_{2k-1})>0$
and $\lambda_{2k} = \overline{\lambda_{2k-1}}$,
we calculate that $f(\lambda_{2k-1}) = 1$ and $f(\lambda_{2k}) = 1$.
\end{proof}

Proposition \ref{Dnuprop} implies that, if we maintain the choice $\vD=0$, any deformation
of the branch points that is analytic in parameter $\epsilon$ will induce an analytic
deformation of the initial values $\nu_k(0,0)$.  This is also true if we increase
the genus by analytically splitting a double point of the Floquet spectrum at $\epsilon=0$ into two
branch points for $\epsilon \ne 0$, since the extra $\nu$-eigenvalue that is unlocked
at the higher genus still belongs to the Dirichlet spectrum as it limits to a double
point when $\epsilon=0$.  It is also true for the isoperiodic deformations described
in \S\ref{isodeforms} that any other double point deforms analytically in $\epsilon$,
and so any locked Dirichlet eigenvalues are analytic in $\epsilon$.

The $\mu$-eigenvalues also deform analytically in $\epsilon$.  For, when $\vD=0$,
the function $f$ satisfies $f(\interchange P) = 1/f(P)$.  Thus,
\[h(\lambda) = f(P) + f(P)^{-1}, \qquad \lambda = \pi(P)\]
is a well-defined meromorphic function on the complex plane with pole
divisor $\pi(D_+ + \infty_+)$.  The zeros of this function are
$\mu_k(0,0)$, so it has the form
\[h(\lambda) = \dfrac{\prod_{k=0}^g (\lambda - \mu_k(0,0))}{r(\lambda)},\]
where $r(\lambda)$ is a polynomial of degree $g$.
At each branch point $h(\lambda)=\pm 2$; the $\nu_k(0,0)$ are zeros of $h(\lambda)-2$, while
the remaining branch points are zeros of $h(\lambda)+2$.  Because the coefficients of
the polynomials in the numerator of $h(\lambda)\pm 2$ must be analytic in $\epsilon$,
it follows that the coefficients of the numerator of $h(\lambda)$ are analytic in $\epsilon$.
Then the $\mu_k(0,0)$ are analytic in $\epsilon$ so long as they are distinct;
however, distinctness holds for sufficiently small $\epsilon$ at each deformation step.

\subsection{Trace Formulas}
The NLS solution $q(x,t)$ is determined by the Dirichlet spectrum and branch points by
the trace formulas%
\footnote{These are adapted from \cite{MOv}, after making the
changes $q \mapsto -2q$, $r\mapsto -2r$, $\lambda \mapsto -\lambda$ which
are necessary to make their version of NLS and its Lax pair coincide with ours.}
\begin{align*}
q(x,t) +r(x,t) &= \sum_{k=0}^g \left(\lambda_{2k} + \lambda_{2k-1} - 2 \mu_k(x,t)\right)\\
q(x,t) -r(x,t) &= -\ri \sum_{k=0}^g \left(\lambda_{2k} + \lambda_{2k-1} -2\nu_k(x,t)\right),
\end{align*}
where we take the specialization $r= -\overline{q}$ for the focusing case.

Given the branch points and the initial values $\mu_k(0,0)$ and $\nu_k(0,0)$, the
values of $\mu_k(x,t)$ and $\nu_k(x,t)$ are determined by ODE systems
derived by Ablowitz and Ma \cite{AMa}:
\begin{align*}
\dfrac{\di \mu_k}{\di x} &= \dfrac{\ri \zeta(\mu_k)\left(c - \sum_{j\ne k} \mu_j\right)}{\prod_{j\ne k} (\mu_j-\mu_k)},\\
\dfrac{\di\mu_k}{\di t} &= 2\mu_k\dfrac{\di\mu_k}{\di x}+
\dfrac{2\ri\zeta(\mu_k)}{\prod_{j\ne k} (\mu_j-\mu_k)}
\left[\left(\dfrac{c}2-\sum_{j=0}^g\mu_j\right)^2 + \left(\dfrac{c}2-\sum_{j=0}^g\nu_j\right)^2
-\sum_{j=0}^g\dfrac{\di \nu_j}{\di x}\right],
\end{align*}
where $c = \sum_{m=1}^{2g+2} \lambda_m$,
and
\begin{align*}
\dfrac{\di \nu_k}{\di x} &= \dfrac{\ri \zeta(\nu_k)\left(c - \sum_{j\ne k} \nu_j\right)}{\prod_{j\ne k} (\nu_j-\nu_k)},\\
\dfrac{\di\nu_k}{\di t} &= 2\nu_k\dfrac{\di\nu_k}{\di x}+
\dfrac{2\ri\zeta(\nu_k)}{\prod_{j\ne k} (\nu_j-\nu_k)}
\left[\left(\dfrac{c}2-\sum_{j=0}^g\mu_j\right)^2 + \left(\dfrac{c}2-\sum_{j=0}^g\nu_j\right)^2
-\sum_{j=0}^g\dfrac{\di \mu_j}{\di x}\right].
\end{align*}

In these systems, the branch points $\lambda_m$ and initial
values $\mu_k(0,0), \nu_k(0,0)$ depend analytically on $\epsilon$.
Our choice $\vD=0$ implies that the initial values $\nu_k(0,0)$ are
located at branch points, where $\zeta=0$.  Thus, the $\nu_k$ are
constant in $x$ and $t$, and are automatically analytic in $\epsilon$.

In the system for the $\mu_k$,
$\zeta(\mu_k)$ is to evaluated along the upper sheet.
(According to \cite{FL}, the $\mu$-eigenvalues travel along the $a$-cycles of the Riemann surface.)
Note that when $\epsilon=0$, the
initial values of one of the $\mu$-eigenvalues is
a double point, the limit of two complex conjugate branch points
that coalesce as $\epsilon\to 0$.
The above characterization of the $\mu_k(0,0)$ as points where $f(P)=-\ri$, coupled with
the fact that $f(\tau P) = - \overline{f(P)}$ ensures that the $\mu_k(0,0)$ lie
along the real axis.  Thus, each $\zeta(\mu_k(0,0))$ will be analytic in $\epsilon$, including
at $\epsilon =0$.


It follows that the $\mu_k(x,t)$ and $\nu_k(x,t)$
are analytic in $\epsilon$ for any $x$ and $t$.
Then the trace formulas imply that $q(x,t)$
is analytic in $\epsilon$.

%% file: isobib.tex
\newcommand{\ditto}{{\leavevmode\vrule height 2pt depth -1.6pt width 23pt\,}}